\documentclass[]{aa} 
\usepackage{graphicx}
\usepackage{txfonts}
\usepackage[breaklinks=true]{hyperref}
\hypersetup{colorlinks=true,citecolor=blue}

\newcommand{\tile}{{\textcolor{black}{{tile}}}}

\newcommand{\pawprint}{{\textcolor{black}{{pawprint}}}}
\newcommand{\pawprints}{{\textcolor{black}{{pawprints}}}}
\newcommand{\stapawprint}{{\textcolor{black}{{stacked pawprint}}}}
\newcommand{\stapawprints}{{\textcolor{black}{{stacked pawprints}}}}
\newcommand{\deeppawprint}{{\textcolor{black}{{deep pawprint}}}}
\newcommand{\deeppawprints}{{\textcolor{black}{{deep pawprints}}}}

\newcommand{\sdi}{{\textcolor{black}{{stacked SDI}}}}
\newcommand{\sdis}{{\textcolor{black}{{stacked SDIs}}}}
\newcommand{\stasdi}{{\textcolor{black}{{deep SDI}}}}
\newcommand{\stasdis}{{\textcolor{black}{{deep SDIs}}}}

\newcommand{\pattern}{Moir\'{e} patterns}

\newcommand{\myemph}[1]{{\textcolor{black}{#1}}}

\newcommand{\revionemz}[1]{{\textcolor{black}{#1}}}
\newcommand{\revitwomz}[1]{{\textcolor{black}{#1}}}
\newcommand{\revile}[1]{{\textcolor{black}{#1}}}

\begin{document} 

\title{Deep PSF photometric catalog of the VVV survey data\thanks{The full data of Table~\ref{tab:catalog_para} are only available in electronic form at the CDS via anonymous ftp to cdsarc.u-strasbg.fr (130.79.128.5)
or via http://cdsweb.u-strasbg.fr/cgi-bin/qcat?J/A+A/....}}
\author{M. Zhang
          \inst{1,2}
          \and
        J. Kainulainen
        \inst{3,1}
    }
            

\institute{Max-Planck-Institut f\"{u}r Astronomie,
              K\"{o}nigstuhl 17, D-69117 Heidelberg, Germany\\
              \email{miaomiao@pmo.ac.cn}
         \and
             Purple Mountain Observatory, and Key Laboratory for Radio Astronomy, Chinese Academy of Sciences, Nanjing 210008, China
         \and
                        Chalmers University of Technology, Department of Space, Earth and Environment, SE-412 93 Gothenburg, Sweden
             }

   \date{}

\abstract
   {The Vista Variables in the V\'ia L\'actea (VVV) survey has performed a multi-epoch near-infrared imaging of the inner Galactic plane. High-fidelity photometric catalogs are needed to utilize the data.}
   {We aim at producing a deep, point spread function (PSF) photometric catalog for the VVV survey $J$-,$~H$-, and $K_s$-band data. Specifically, we aim to take advantage of all the epochs of the survey to reach high limiting magnitudes.}
   {We developed an automatic PSF-fitting pipeline based on the DaoPHOT algorithm and performed photometry on the stacked VVV images in $J,~H$, and $K_s$ bands.}
   {We present a PSF photometric catalog \revionemz{in the Vega system} that contains about 926 million sources in the $J,~H$, and $K_s$ filters. About \revionemz{10\%} of the  sources are flagged as possible spurious detections. The 5$\sigma$ limiting magnitudes of the sources with high reliability are \revitwomz{about} 20.8, 19.5, and 18.7 mag in the $J,~H$, and $K_s$ bands, respectively\revitwomz{, depending on the local crowding condition}. Our photometric catalog reaches on average about one magnitude deeper than the previously released PSF DoPHOT photometric catalog and includes less spurious detections. There are significant differences in the brightnesses of faint sources between our catalog and the previously released one. The likely origin of these differences is in the different photometric algorithms that are used; it is not straightforward to assess which catalog is more accurate in different situations. Our new catalog is beneficial especially for science goals that require high limiting magnitudes\revitwomz{; our catalog reaches such high maginitudes in fields that have a relatively uniform source number density. Overall, the limiting magnitudes and completeness are different in fields with different crowding conditions.}} 
    {}
   \maketitle


\section{Introduction}


The Vista Variables in the V\'ia L\'actea (VVV, \citealt{vvv}) is an ESO public survey that has imaged approximately 562 square degrees of the inner Galactic plane in near-infrared with the VIRCAM instrument \citep[VISTA InfraRed CAMera;][]{vircam1,vircam2} on the VISTA telescope \revionemz{\citep{vista-ref}}. The survey covers five bands, namely $Z, Y, J, H$, and $K_s$, and extends over a time period of five years \citep{vvv, vvvdr1}. While the main science driver of the survey is the variability of stars in the Galactic bulge and the disk, the wealth of data it produces can be used for a wide spectrum of purposes, ranging from revealing Galactic structures \citep[e.g.,][]{gon11,minniti11,wegg13,minniti14,simion17,gon18} to census of star clusters \citep[e.g.,][]{bor11,moni11,chen12,barba15,minn17}, young stellar objects \citep[e.g.,][]{mattern18, zhang19}, variable stars \citep[e.g.,][]{ang14,ag15,cha15,gran15,nav16,con17}, brown dwarfs \citep{beam13,smith15}, high-proper-motion stars \citep{kur17}, and microlensing events \citep{minni15}. 

The VVV images contain hundreds of millions of stars above the detection limits of the survey. Obtaining photometric measurements for all these sources poses a fundamental challenge for data processing and analysis because  performing accurate photometry in the crowded conditions of the Galactic plane is very difficult. Various approaches to confronting this challenge exist; the choice between them is not trivial and depends on the qualities desired for the data (see, e.g., discussion in Sections \ref{sect:psfphot} and \ref{sect:photcali}). This basic problem is crucially enhanced by the sheer amount of data in the VVV survey; handling them requires fully automated, parallelized procedures for all relevant tasks. \revile{Indeed, various photometric techniques have been applied to the VVV survey data in the past \citep{vvvdr1,skzpipeline,vvvpsf,surot19b,surot19a}.} Recently, the VVV science team has published the first point spread function (PSF) photometric catalog for the whole VVV survey \citep{vvvpsf}. The catalog contains approximately 846 million sources with detections in at least three filter bands of $Z, Y, J, H$, and $K_s$;  detections in the $J$, $H$, and $K_s$ bands have been made for about
534 million of these sources (see Section \ref{sect:data} for further details). 

In this paper, we develop a new PSF photometric pipeline and apply it to the VVV survey $J$-, $H$-, and $K_{s}$-band data. Our work is motivated by the interest to reach deeper limiting magnitudes than those of the catalog by \citet{vvvpsf}. This interest is driven by our goal to perform dust extinction mapping in the Galactic plane in the future \citep[e.g.,][]{kainulainen11alves, kainulainen13tan, butler14darkest-shadows, kainulainen-snake, mattern18}; for that purpose, detecting as many sources as possible is desired. We show that this can indeed be achieved by exploiting different data reduction and photometry schemes: our final photometric catalog reaches on average roughly one magnitude deeper than the catalog of \citet{vvvpsf}. Here, we describe the automated pipeline that performs the photometry, present its results, and release the catalog to the community.

\section{VVV survey data}\label{sect:data}



Here, we use the near-infrared imaging data of the VVV survey. The VVV survey covers the Galactic bulge (-10 $\leq$ $l$ $\leq$ 10, -10 $\leq$ $b$ $\leq$ 5) and part of the adjacent Galactic plane (-65 $\leq$ $l$ $\leq$ -10, -2 $\leq$ $b$ $\leq$ 2). Below, we describe the data to the degree required to understand the implementation of our data reduction and photometry procedures (described in Section \ref{sec:photometry}). For the detailed description of the survey and the data, we refer to the comprehensive survey papers cited in the text below.

\begin{figure}
    \centering
    \includegraphics[width=1.0\linewidth]{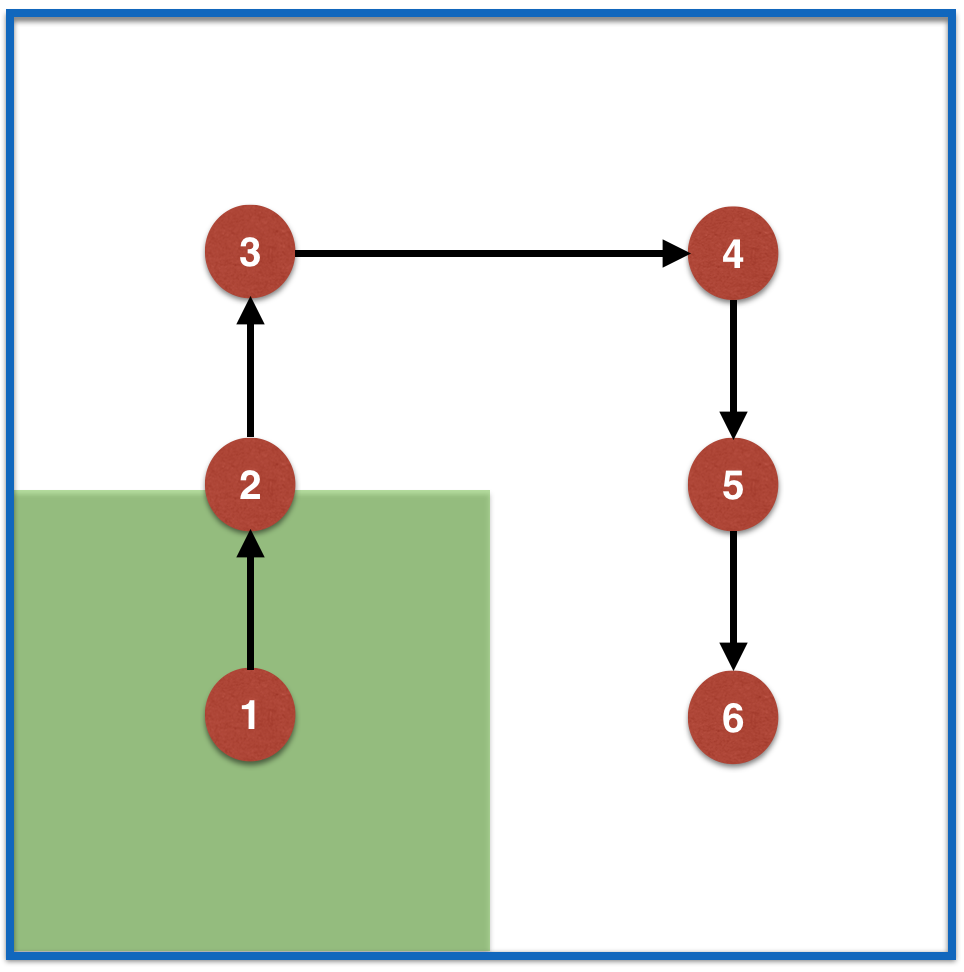}
    \caption{\revitwomz{Sketch of "Tile6n" observing pattern for one VISTA detector. The six dithering pointings are labeled with the red filled circles. The numerals marked on red circles and the arrows show the dithering sequence. The filled green box represents the field of view of a detector and the solid blue square marks the total area covered by six dithering images.}}
    \label{fig:tile6npattern}
\end{figure}

The survey instrument, VIRCAM, is equipped with 16 \revionemz{detectors} 2048 $\times$ 2048 pixels$^2$ in size, with a pixel scale of \revionemz{0.339$\arcsec$}. The \revionemz{detectors} are arranged in a 4 $\times$ 4 array with gaps along the X and Y axes. A single pointing with the detector array is called a \pawprint\ and \revionemz{consists of 16 single detector images (SDIs), covering $\sim$0.595} deg$^2$ on the sky. As with all VISTA observations, the basic observational unit is the observation block (OB\footnote{see \url{https://www.eso.org/sci/observing/phase2/MANUAL/p2ppman_v8.pdf} for details.}). \revitwomz{There are two types of OB in the VVV survey: multi-filter, single-epoch OBs ($JHK_s$ and $ZY$ OBs), and $K_s$-only variability-monitoring OBs.} The VVV OBs use the "Tile6n+Jitter2u" observing pattern, which means there are six pawprint positions and two small offsets at each pawprint position. \revitwomz{Figure~\ref{fig:tile6npattern} shows the "Tile6n" dithering pattern for one VISTA detector.} Two exposures at each pawprint position can be combined to construct a \stapawprint. Six \stapawprints~can be combined to construct a \tile~that covers a contiguous field of \revionemz{$\sim$1.687} deg$^2$. \revionemz{This value is slightly larger than the average size of a VVV tile ($\sim$1.64 deg$^2$) because the adjacent VVV tiles share some overlaps.} 
More information about the instruments and observing strategy can be found in \revionemz{\citet{vista-ref}}, \citet{vvv}, and \citet{vvvdr1}. Altogether, the coverage of the VVV survey consists of 348 tiles, including 196 tiles in the bulge (tile names start with "b"), and 152 tiles in the disk (tile names start with "d"). All tile centers and numbers can be found in \citet{vvvdr1}. 

\begin{table*}
\small
\caption{\revitwomz{Exposure times of VVV OBs}}
\label{tab:exptime}
\centering
\begin{tabular}{cccccccc}
\hline\hline
Area & OB type & Filter & DIT\tablefootmark{a} & NDIT\tablefootmark{b} & $\textrm{Exp}_{\textrm{p}}$\tablefootmark{c} & $\textrm{Exp}_{\textrm{sp}}$\tablefootmark{d} & Median exp. time per pixel \\
 & & & (s) & & (s) & (s) &  of single epoch OB (s)\\
\hline 
Bulge & multi-filter, single-epoch & $J$ & 6 & 2 & 12 & 24 & 48\\ 
Bulge & multi-filter, single-epoch & $H, K_s$ & 4 & 1 & 4 & 8 & 16\\
Bulge & variability monitoring & $K_s$ & 4 & 1 & 4 & 8 & 16\\
Disk & multi-filter, single-epoch & $J, H, K_s$ & 10 & 2 & 20 & 40 & 80\\
Disk & variability monitoring & $K_s$ & 4 & 1 & 4 & 8 & 16\\
\hline                                   
\end{tabular}
\tablefoot{
\tablefoottext{a}{Detector integration time}
\tablefoottext{b}{number of DITs}
\tablefoottext{c}{exposure time per \pawprint}
\tablefoottext{d}{exposure time per \stapawprint}
}
\end{table*}

\revitwomz{Table~\ref{tab:exptime} shows the exposure time of VVV OBs. Each \pawprint~is a co-addition of n exposures and therefore the exposure time per \pawprint~($\textrm{Exp}_{\textrm{p}}$) is equal to the DIT (detector integration time) multiplied by the NDIT (number of detector integration time). In the bulge area, the Exp$_{\textrm{p}}$ is 2$\times$6s, 4s, and 4s for $J$, $H$, and $K_s$, respectively. In the disk area, the Exp$_{\textrm{p}}$~is 2$\times$10s for $J$, $H$, and $K_s$ for the multi-filter, single-epoch OBs, and 4s for the $K_s$-only variability-monitoring OBs. The \stapawprints~are constructed by combining two small-offset jitter \pawprints~and thus the exposure time per \stapawprint~(Exp$_{\textrm{sp}}$) is 2~$\times$ Exp$_{\textrm{p}}$. In one epoch, six \stapawprints~can be combined to obtain a tile image. However, considering the large offsets among six \stapawprints, \revile{each pixel of the tile image is covered by approximately two or more} \stapawprints~and thus the median exposure time per pixel in the tile image is about 2~$\times$ Exp$_{\textrm{sp}} = 4 \times \textrm{Exp}_{\textrm{p}}$~\citep[see also Table 3 of][]{vvvdr1}. The \stapawprints~of n epoch observations can also be combined to construct a \deeppawprint~and the median exposure time per pixel in each \deeppawprint~is about $4 \times n \times \textrm{Exp}_{\textrm{p}}$.}

The VVV data have been processed using VISTA data flow system (VDFS) pipeline at the Cambridge Astronomical Survey Unit (CASU\footnote{\url{http://casu.ast.cam.ac.uk/vistasp/}}) (\revionemz{\citealt{irwin04,cross09}}; \citealt{lewis10,vvvdr1}). The process of VVV data reduction and one approach to calibration and photometry can be found in \citet{vvvdr1} \revionemz{and \citet{vista-cali}}. 
Here we only summarize the main data-reduction process. First, the dark current is subtracted from each raw image and then the linearity is corrected for every detector. The flat-field correction is also applied by dividing by the twilight flats. Second, the sky background model is constructed with the exposures in a given filter within several concatenated OBs. Considering the variability of the near-infrared sky, at least 24 input \pawprints~closest in time (within $\sim$30 min) are combined to produce the background sky model that is then subtracted from each input \pawprint. Finally the jitter stacking is conducted to construct the \stapawprints~after correcting the stripe pattern introduced by the detector readout electronics.

We use the $JHK_s$ \myemph{images and source catalogs} from Data Release 4 (DR4) provided by the VISTA Science Archive (VSA\footnote{\url{http://horus.roe.ac.uk/vsa/}}) \citep{hambly08,cross12}. 
We note that VSA DR4 \myemph{includes one photometric catalog based on aperture photometry.} 
The 5$\sigma$ limiting magnitude of the catalog is $K_s\sim$17$-$18\,mag for most tiles \citep{vvvdr1}, and it drops to $K_s \sim$15$-$16\,mag in the crowded fields close to the Galactic center \citep{vvvdr1}.
Recently, \citet{vvvpsf} performed PSF photometry on the \myemph{multi-epoch} VVV \myemph{images} 
using DoPHOT \citep{dophot0,dophot} \myemph{and used the average fluxes over all epochs to derive the brightnesses of the detected sources}. 
The 5$\sigma$ limiting magnitude of the catalog is $K_s\sim$18 mag in the VVV disk, $K_s\sim$17.5 mag in the VVV innermost bulge, and $K_s\sim$18 mag in the VVV outermost bulge. \myemph{This DoPHOT PSF photometric catalog is also publicly available through the VSA DR4.}

\section{Methods: PSF photometry}
\label{sec:photometry}

In this section, we first justify our choices for the photometric techniques (Section \ref{subsec:daophot}) and then describe the details of our PSF photometry procedure (Section \ref{subsec:psf-pipeline}). 
The resulting catalog is presented in Section \ref{sect:psf-catalog}. To avoid ambiguity, we introduce the nomenclature used.
\revitwomz{The multi-epoch \stapawprints~are combined to construct the deep images. We refer to them as \deeppawprints. Meanwhile, }
we refer to a SDI from a \stapawprint~as a \revitwomz{\sdi}. We \revitwomz{combine} the \sdis~of the same detector from all multi-epoch \stapawprints~together to obtain the \revitwomz{\stasdis}~(see Section~\ref{sect:stacking}).

\subsection{DaoPHOT versus DoPHOT}\label{sect:psfphot}
\label{subsec:daophot}

Aperture photometry is known to be relatively inaccurate for crowded fields; this is definitely so for the sensitive VVV data, especially at the latitudes close to the Galactic plane and in the Galactic bulge. To achieve accurate photometry in the areas of high stellar density, and to reach deeper limiting magnitude than with the aperture photometry, we perform PSF photometry of the VVV data. 

We considered two commonly used PSF-fitting programs, namely DoPHOT and DaoPHOT \citep{daophot}. DoPHOT uses an analytic function to model the stellar PSF. It is unable to fit more than two stars simultaneously \revionemz{although DoPHOT considers double stars}. The main advantage of DoPHOT is that it runs very quickly and in an entirely automated manner. DaoPHOT combines both the analytic and empirical methods to model the stellar PSF. It fits the central part of a bright star with an analytic function and then uses look-up tables to do corrections from the analytic function to match the observed stellar profile. Once a PSF model is derived, DaoPHOT tries to fit many stars simultaneously. To do that, DaoPHOT first separates all stars into groups and then performs PSF-fitting for all stars in a group simultaneously. The disadvantage of DaoPHOT is that it requires careful configuration of input parameters and is relatively slow.

\citet{skzpipeline} compared the DaoPHOT photometric results with DoPHOT and aperture photometry using the VVV survey data of the globular cluster M22. These latter authors found that DaoPHOT can return a higher detection rate of faint sources and therefore goes deeper than DoPHOT and aperture photometry. Because we wish to include as many sources as possible in the extinction mapping, we decided to use DaoPHOT to construct a new, deep PSF photometric catalog of the VVV data.


\subsection{Automatic PSF-fitting pipeline}
\label{subsec:psf-pipeline}

\begin{center}
\begin{figure}
\includegraphics[width=1.0\linewidth]{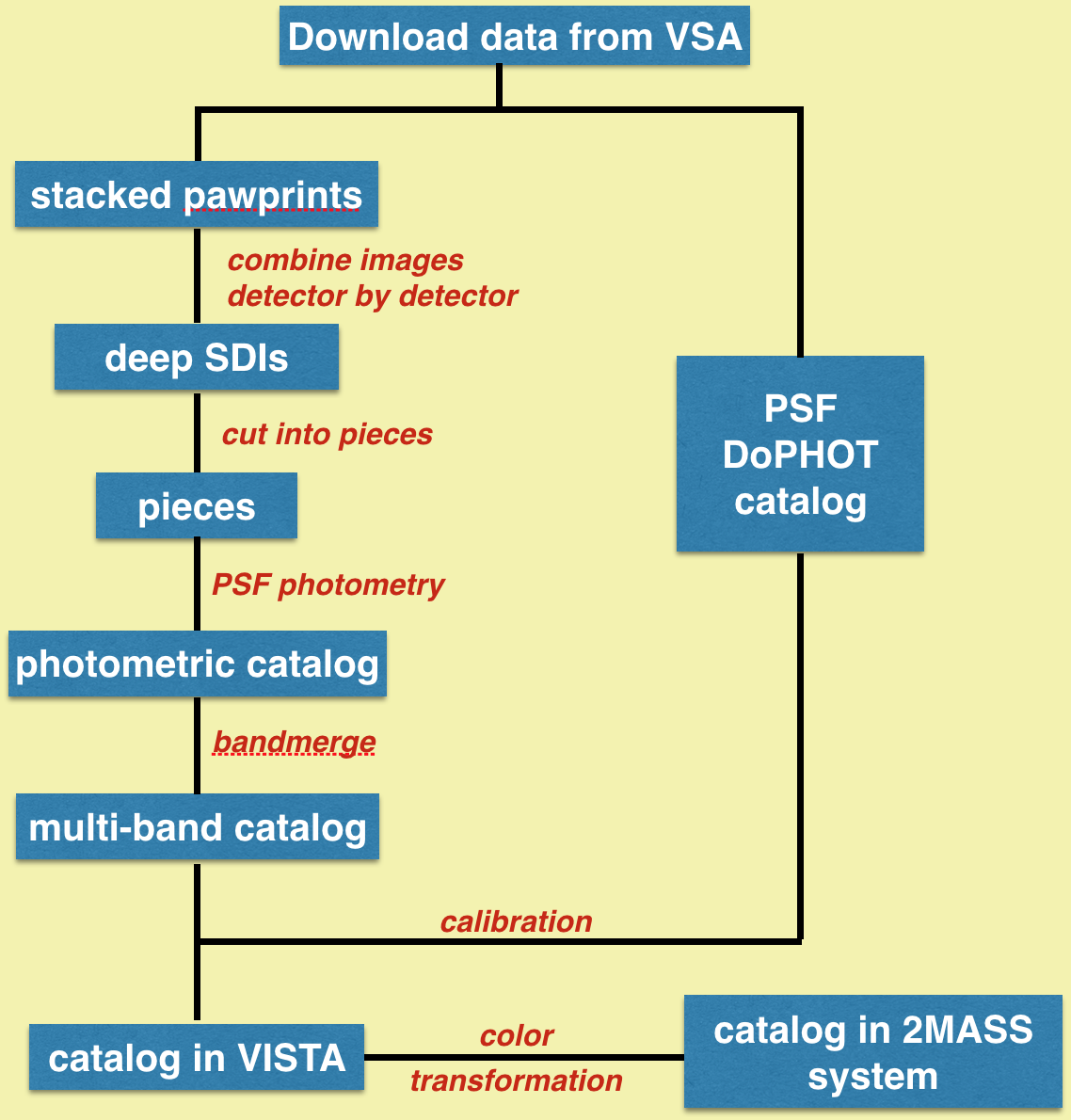}
\caption{Flow chart of our automatic PSF-fitting pipeline that is used to do the PSF photometry on the VVV survey imaging data.}
\label{flowchart-pipeline}
\end{figure}
\end{center}

We developed an automatic pipeline to perform PSF photometry on the VVV survey imaging data. The pipeline uses the DaoPHOT algorithm and is mainly written in IDL and Python. The pipeline is adapted to run in multi-core mode, which  significantly reduces the required CPU time. Figure~\ref{flowchart-pipeline} shows the flow chart of the pipeline. 
We give the detailed description of the pipeline in the following sections (Sections \ref{sect:stacking}-\ref{subsubsec:removal}). 

\subsubsection{Image stacking}\label{sect:stacking}

To reach as deep a photometry as possible, we stack together multi-epoch images for 
each filter, that is, $J$, $H$, and $K_s$. In principle, there are 
two options for stacking, one starting with \tile~images and the other starting with \stapawprints. 
Considering that each \revionemz{detector} of the detector array potentially has a different PSF and that seeing conditions can vary between images, the PSF is expected to vary across a tile. This makes it difficult to model the PSF with a universal function. To avoid the problems arising from the PSF variation, we decided to work with the \stapawprints~(as also suggested by \citealt{skzpipeline}). 

\begin{center}
    \begin{figure*}
        \centering
        \includegraphics[width=1.0\linewidth]{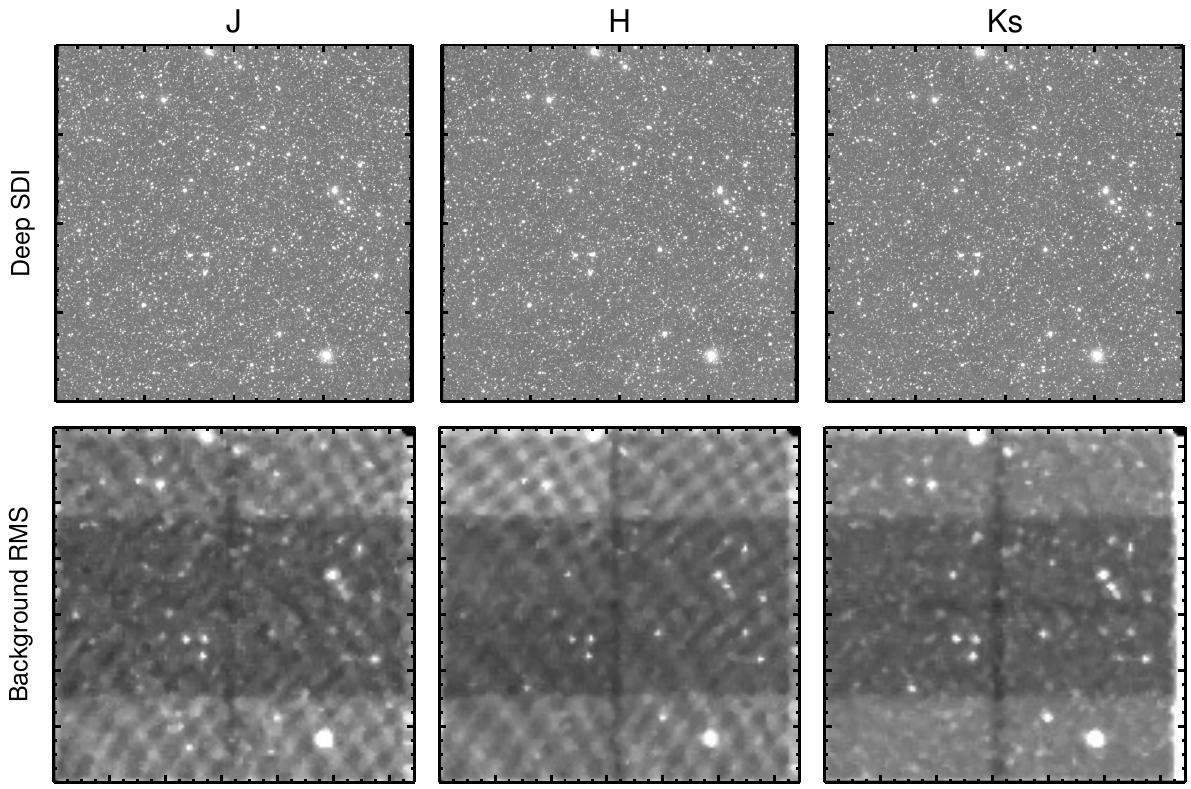}
        \caption{Deep SDIs (top panels) and the corresponding rms noise maps (bottom panels) for one \revionemz{detector (\revitwomz{\#11})} of the "b214" tile in $J$ (left), $H$ (middle), and $K_s$ (right) bands.}
        \label{fig:pattern}
    \end{figure*}
\end{center}

In practice, we download \revitwomz{"good" (deprecated=0)\footnote{This parameter marks the status of the image records in the VSA database. All data that have deprecated=0 are nominally good data. The details can be found in \url{http://horus.roe.ac.uk/vsa/www/gloss_d.html\#multiframe_deprecated}}} \stapawprints~of multiple epochs for each tile region from VSA. For the $n$ epoch observations in one tile region for one filter, there are usually 6$\times n$ \stapawprints. We split 
each \stapawprint~into 16 \sdis. In each filter band, about 6$\times n$ \sdis~corresponding to the same detector are checked and then \revitwomz{combined}  using the SWARP software \citep{swarp} after excluding the \sdis~obtained in bad weather conditions. The detailed process is as follows. First, we calculate a common center based on the centers of the \sdis~that correspond to the same detector. We then reproject and resample each \sdi~onto a common reference frame defined by the common center; this is done using the bilinear interpolation based on the astrometric WCS information of the \stapawprints~provided by CASU. We also subtract a constant background from each \sdi~during the above resampling. Finally, the original resampled \sdis~are co-added to the \stasdis~with SWARP. We note that the median WCS rms of the CASU astrometric solution is $\sim$70\,mas \citep{saito12}. The position uncertainty introduced by the SWARP resampling process is $\sim$0.03 pixels, corresponding to $\sim$10\,mas. Thus, the position accuracy of the \stasdis~is dominated by the CASU astrometric solution. We finally obtain 16 \stasdis~in each filter band of each tile region. As an example, Fig.~\ref{fig:pattern} (top panels) shows the \stasdis~for one detector of the tile "b214" in $J$, $H$, and $K_s$ bands. To avoid the non-negligible field distortion across a tile image \citep{vista-ref,irwin04}, we did not mosaic these 16 \stasdis~to a stacked tile.

We note that the CASU data-reduction pipeline used fast bilinear interpolation to obtain the \stapawprints. We also use bilinear interpolation to stack the multi-epoch \revionemz{\sdis}. However, the bilinear interpolation can introduce spatially correlated noise in the final images (as pointed out by \citealt{vision}). Such a phenomenon is illustrated in Fig.~\ref{fig:pattern} (bottom panels), which shows the background rms maps of \revionemz{\stasdis} for one \revionemz{detector} of tile "b214". We can see obvious Moir\'{e} patterns in $J$ and $H$ background rms maps. 
Unfortunately, the spatially correlated noise introduced by the bilinear interpolation can also affect photometry. It can produce Moir\'{e} patterns in the color space, which would further transfer into systematic patterns in extinction maps (the extinction mapping is based on colors of stars). Our solution is to lessen the effect of these patterns during the photometric calibration process (explained in Section \ref{sect:photcali}). While this produces satisfactory results for our goal, complete elimination of the spatially correlated noise would require a re-processing of all VVV survey raw data using more suitable interpolation methods (e.g., a higher-order resampling kernel as suggested by \citealt{vision}); this is clearly beyond the scope of our paper.  

\subsubsection{Source detection and photometry}

\revionemz{The source detection and photometry is performed on the \stasdis.}
To better handle the large amount of data and parallelise the photometry, we split each \revionemz{\stasdi} into pieces of $\sim$1000 $\times$ 1000 pixels. The PSF photometry is performed on each piece with PyRAF\footnote{\url{http://www.stsci.edu/institute/software_hardware/pyraf}} that is a Python front for running IRAF\footnote{\url{http://iraf.noao.edu/}} tasks. The source detection is performed using the DAOFIND task with the signal-to-noise ratio threshold of $S/N > $~3. The PSF function is modeled with the PSF task of DaoPHOT. Finally, we use ALLSTAR task to obtain the photometric results. 

\begin{center}
\begin{figure}
\includegraphics[width=1.0\linewidth]{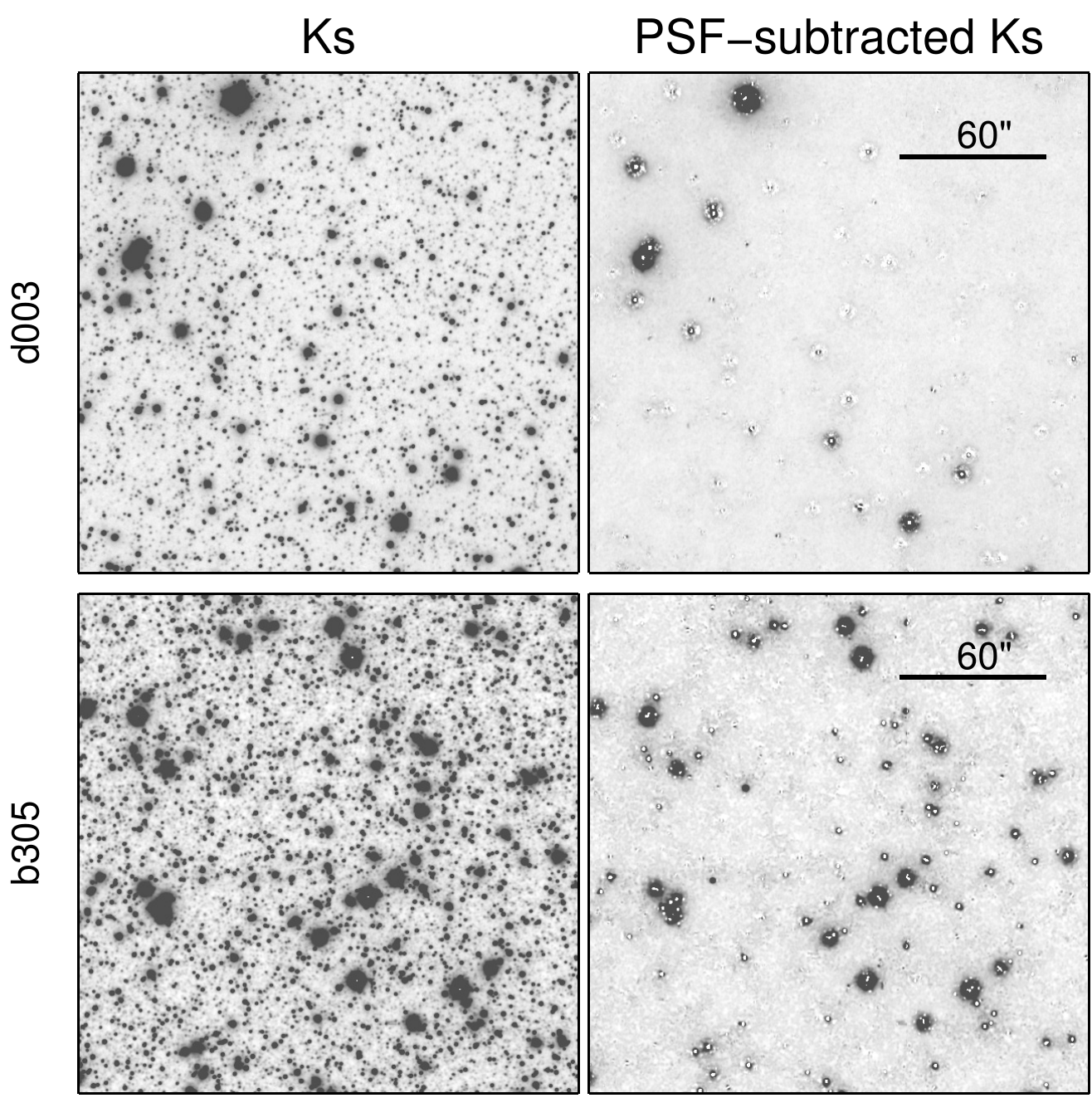}
\caption{Partial \revionemz{\stasdis~\revitwomz{($\sim$3\arcmin$\times$3\arcmin)} at $K_s$} of tile "d003" (\textit{top panels}) and "b305" (\textit{bottom panels}) before (\textit{left panels}) and after (\textit{right panels}) PSF subtraction.}
\label{psfsubshow}
\end{figure}
\end{center}

To illustrate the quality of the process, Fig.~\ref{psfsubshow} shows two partial regions of the tiles "d003" and "b305" before and after PSF subtraction in $K_s$ band. The tile "d003" is located in the VVV disk region and represents a relatively uncrowded field. The tile "b305" represents a highly crowded field close to the Galactic center. 
The example shows that most stars have been well fitted and subtracted from the stacked images. The residuals around bright stars in the PSF-subtracted images are mainly due to the saturation and nonlinearity effects.

For each piece of the \revionemz{\stasdis}, an instrumental PSF-fitting photometric catalog is obtained from the output of the DaoPHOT ALLSTAR task. We apply the astrometric WCS information \revionemz{recorded in each corresponding \stasdi~(see Section~\ref{sect:stacking})}. We create a single-band catalog by cross-matching the WCS coordinates of the sources in all pieces of the 16 \revionemz{detectors} of a VVV tile. This is done using the STILTS package \citep{stilts} with a tolerance of 0\farcs5. We then merge the single-band catalogs using STILTS with a tolerance of 1\arcsec~\revitwomz{\citep{vvvdr1}. For band-merging about 93-98\% (with a mean value of $\sim$96\%) of the sources are matched within 0\farcs5.} 

\subsubsection{Photometric calibration}
\label{sect:photcali}

\begin{figure*}
    \centering
    \includegraphics[width=1.0\linewidth]{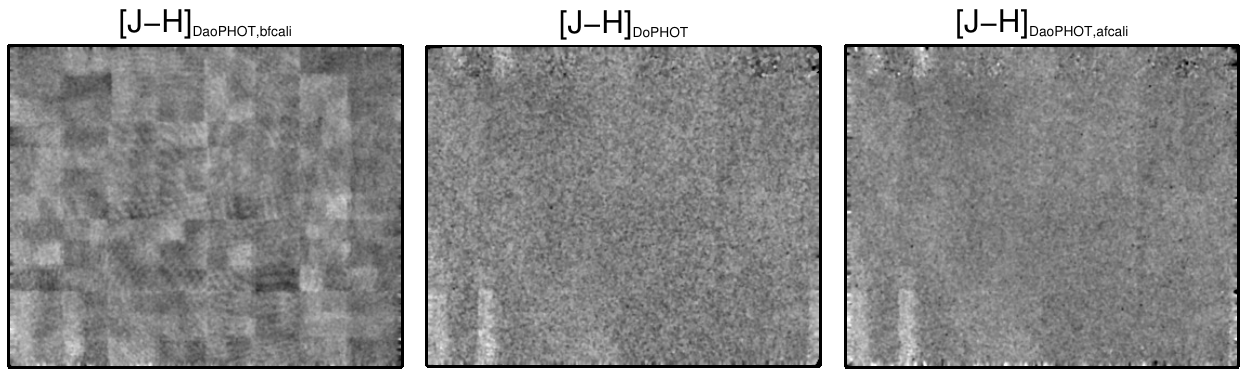}
    \caption{$J-H$ color images of sources in tile "b214" of our DaoPHOT catalog before the photometric calibration (left), of the released DoPHOT catalog \citep[middle,][]{vvvpsf}, and of our DaoPHOT catalog after the photometric calibration (right). The images are produced by smoothing the source $J-H$ colors with a Gaussian kernel (FWHM=30\arcsec).}
    \label{fig:jhcolorimage}
\end{figure*}

The absolute photometric calibration was obtained by a comparison of the instrumental magnitudes of relatively isolated, unsaturated bright sources with their counterparts in the released DoPHOT PSF photometric catalog \citep{vvvpsf}. As mentioned in Sect~\ref{sect:stacking}, our DaoPHOT photometry is affected by the noise patterns introduced by bilinear interpolation. Figure~\ref{fig:jhcolorimage} shows $J-H$ color maps made from our DaoPHOT catalog before the photometric calibration and from the released DoPHOT catalog. We can see obvious \pattern~in the $J-H$ color map of our DaoPHOT photometry. There are no \pattern~in the DoPHOT-based color map. This can be understood by considering the differences in our data treatment. \citet{vvvpsf} performed DoPHOT PSF photometry on every \revionemz{\sdi}, which are also affected by the noise patterns. However, using the photometric uncertainties as weights, these latter authors averaged the brightnesses of the sources detected in more than one \revionemz{\sdi}. This averaging process dealt with the noise patterns. Since the noise patterns are not present in the DoPHOT catalog, we decided to use it to calibrate our DaoPHOT photometry and remove, to a certain extent, the \pattern. The following general transformation equations were used in this process:
\begin{align}
\begin{split}
    J_{\textrm{DoP}}-J_{\textrm{DaoP}}&=z_{J}+c_{JH}(J_{\textrm{DaoP}}-H_{\textrm{DaoP}})\\
    &+c_{HK_s}(H_{\textrm{DaoP}}-K_{s\textrm{DaoP}})+c_{JK_s}(J_{\textrm{DaoP}}-K_{s\textrm{DaoP}})\\
    &+b_{J}P_J(x,y)+b_{JH}P_{JH}(x,y)\\
    &+b_{HK_s}P_{HK_s}(x,y)+b_{JK_s}P_{JK_s}(x,y),
    \end{split}
    \label{eqj}
\end{align}

\begin{align}
    \begin{split}
    H_{\textrm{DoP}}-H_{\textrm{DaoP}}&=z_{H}+c_{JH}(J_{\textrm{DaoP}}-H_{\textrm{DaoP}})\\
    &+c_{HK_s}(H_{\textrm{DaoP}}-K_{s\textrm{DaoP}})+c_{JK_s}(J_{\textrm{DaoP}}-K_{s\textrm{DaoP}})\\
    &+b_{H}P_H(x,y)+b_{JH}P_{JH}(x,y)\\
    &+b_{HK_s}P_{HK_s}(x,y)+b_{JK_s}P_{JK_s}(x,y),
    \end{split}
    \label{eqh}
\end{align}

\begin{align}
    \begin{split}
    K_{s\textrm{DoP}}-K_{s\textrm{DaoP}}&=z_{K_s}+c_{JH}(J_{\textrm{DaoP}}-H_{\textrm{DaoP}})\\
    &+c_{HK_s}(H_{\textrm{DaoP}}-K_{s\textrm{DaoP}})+c_{JK_s}(J_{\textrm{DaoP}}-K_{s\textrm{DaoP}})\\
    &+b_{K_s}P_{K_s}(x,y)+b_{JH}P_{JH}(x,y)\\
    &+b_{HK_s}P_{HK_s}(x,y)+b_{JK_s}P_{JK_s}(x,y).
    \end{split}
    \label{eqk}
\end{align}
In these equations, $z, c,$ and $b$ are the zeropoints, coefficients of color terms, and coefficients of functions that are related to the positions, respectively. $P(x,y)$ describes the magnitude or color difference between the DoPHOT and DaoPHOT photometries as a function of the image position and therefore also traces the \pattern. 
$P(x,y)$ is constructed from a smoothed differential between the source magnitudes, or colors, in the DoPHOT and DaoPHOT photometries. For example, $P_{J}(x,y)$ is obtained by smoothing the [$J_{\textrm{DoP}}-J_{\textrm{DaoP}}$] of sources using a Gaussian kernel with the FWHM of 10\arcsec-30\arcsec. The value of the FWHM depends on the source number density of the field; we require at least five sources within each beam to facilitate the smoothing.  

\begin{figure*}
    \centering
    \includegraphics[width=1.0\linewidth]{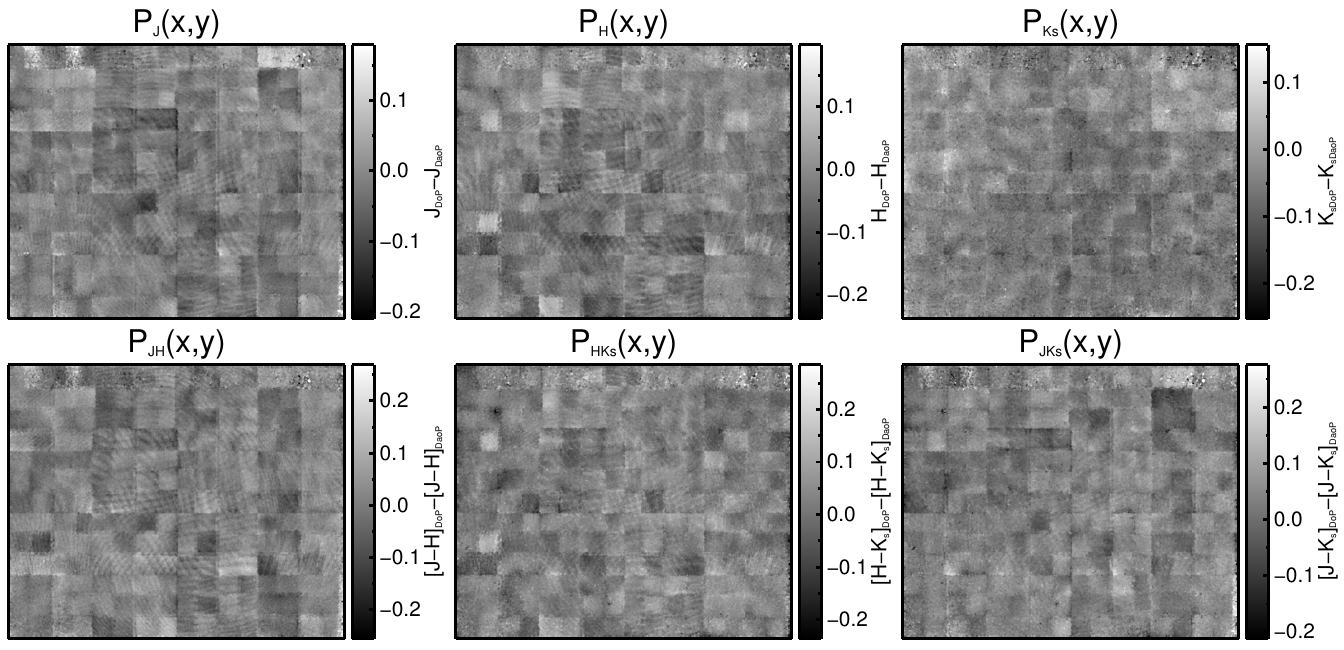}
    \caption{Function $P(x,y)$ constructed in tile "b214" in the first iteration.}
    \label{fig:pxy}
\end{figure*}

In practice, the calibration proceeds as follows for each tile. 
\begin{itemize}
\item[1)] We crossmatch our DaoPHOT photometry with DoPHOT catalog using a tolerance of 0.34\arcsec~($\sim$1 pixel). Considering the standard deviations of $J_{\textrm{DoP}}-J_{\textrm{DaoP}}$, $H_{\textrm{DoP}}-H_{\textrm{DaoP}}$, and $K_{s\textrm{DoP}}-K_{s\textrm{DaoP}}$ ($\sigma_{J}, \sigma_{H}, \sigma_{K_s}$), we also exclude sources with $J_{\textrm{DoP}}-J_{\textrm{DaoP}}>3\sigma_{J}$ or $H_{\textrm{DoP}}-H_{\textrm{DaoP}}>3\sigma_{H}$, or $K_{s\textrm{DoP}}-K_{s\textrm{DaoP}}>3\sigma_{K_s}$.  
\item[2)] We construct $P(x,y)$ with the matched sources. Figure~\ref{fig:pxy} shows the $P(x,y)$ constructed in tile "b214". 
\item[3)] We select the matched unsaturated sources with uncertainties of $<0.05$ at $J, H,$ and $K_s$. Using these high-quality matched sources, a linear regression is performed to solve the transformation equations~\ref{eqj}, \ref{eqh}, and~\ref{eqk}.  
\item[4)] We apply the obtained zeropoints and coefficients to our DaoPHOT catalog to obtain the calibrated photometric results. \end{itemize}
We repeat the above four steps 5-10 times as long as the \pattern\ keep decreasing and then obtain the final photometry. 
To illustrate the result, Fig.~\ref{fig:jhcolorimage} shows the $J-H$ color distribution of sources in our DaoPHOT catalog after the iterative photometric calibration. The \pattern~in the $J-H$ color space have been \revile{almost completely removed}.

\begin{figure*}
    \centering
    \includegraphics[width=1.0\linewidth]{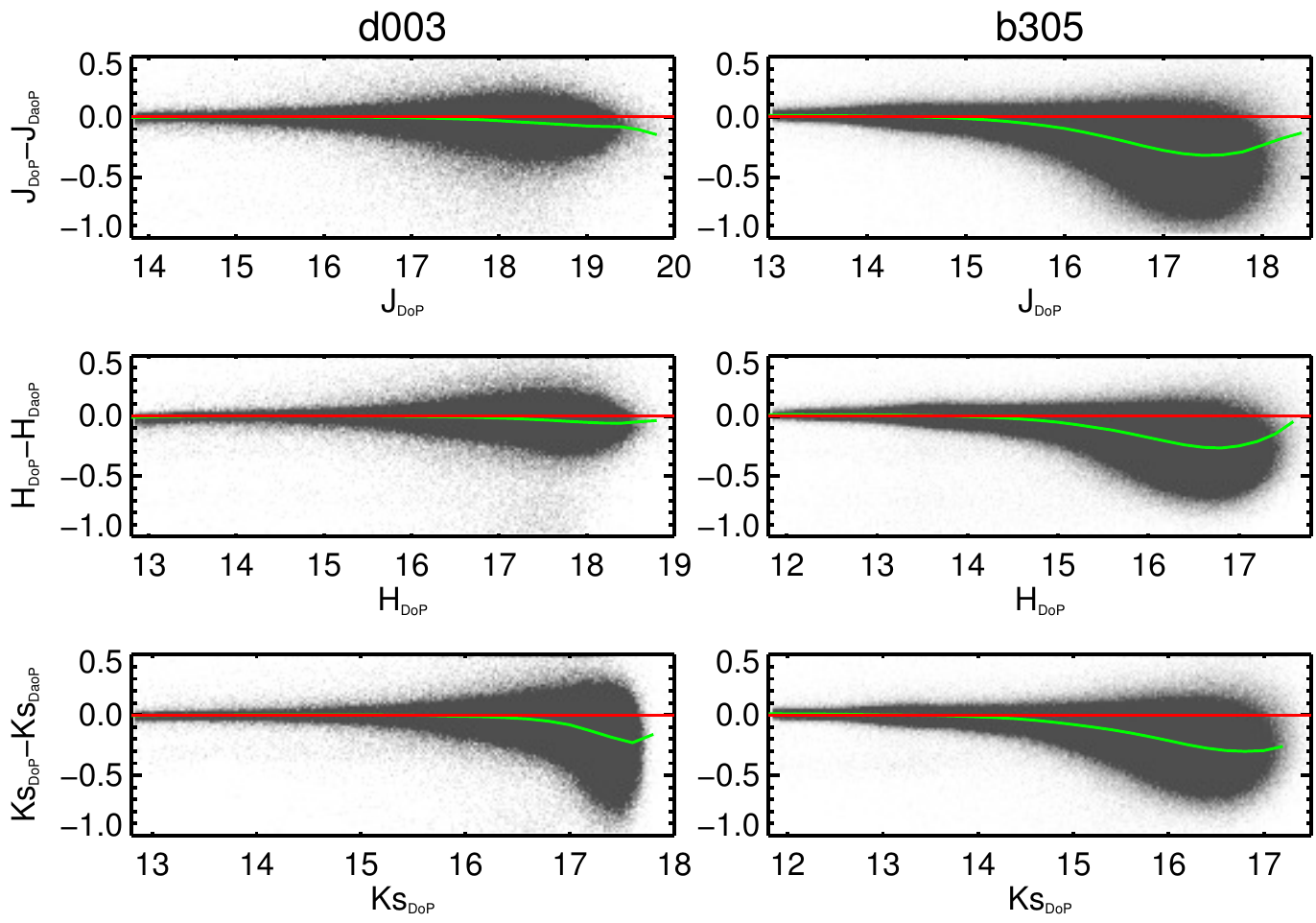}
    \caption{Released DoPHOT photometry vs. our DaoPHOT photometry in tiles "d003" (left panels) and "b305" (right panels). The backgrounds are source number density maps. The green curves show moving medians with a box width of 0.2 mag while the red lines mark the zero difference value.}
    \label{fig:dovsdao}
\end{figure*}

Figure~\ref{fig:dovsdao} shows a comparison between DoPHOT photometry and our calibrated DaoPHOT photometry for sources in the tiles "d003" and "b305". They match at the bright end, but there are systematic differences ($\sim$0.1-0.3 mag) at the faint end. DoPHOT tends to yield brighter photometry than DaoPHOT for the faint sources. Considering that the average source number density in the bulge tile "b305" is higher than that in the disk tile "d003", it seems that this systematic difference is larger for more crowded fields. This systematic difference between DoPHOT and DaoPHOT for faint stars has been previously seen in studies of stellar clusters \citep{friel91,janes93,hill98}. In particular, \citet{friel91} compared DoPHOT and DaoPHOT in a cluster field and an uncrowded field. They found that the systematic difference between DoPHOT and DaoPHOT is small in the uncrowded field. However, in the cluster field the difference is a function of position relative to the center of cluster, increasing towards the cluster center. This systematic difference could be due to the different methods of sky background estimation used in DoPHOT and DaoPHOT. DaoPHOT estimates sky flux based on the PSF-subtracted images while DoPHOT measures the sky flux in the surroundings of stars without considering the PSF-fitting process. This usually results in higher sky estimations by DaoPHOT than by DoPHOT \citep{friel91,dophot,hill98}. Overall, the two photometry packages adopt different philosophies and use different algorithms. As a result, it is difficult to judge which one is "better" without a direct calibration of the faint stars. Adding to this problem, experiments on simulated frames with artificial stars have found that neither DoPHOT nor DaoPHOT can recover the true flux of faint stars in crowded fields \citep{dophot}. Therefore, we here note that there are systematic differences between the released DoPHOT catalog and our DaoPHOT photometry for the faint sources, even though we use the bright sources of the DoPHOT catalog  to calibrate our DaoPHOT catalog. In Section~\ref{sect:colordiff}, we revisit this issue and quantify the systematic difference between the DaoPHOT and DoPHOT photometric results through a detailed comparison.

\begin{center}
\begin{figure*}
\includegraphics[width=1.0\linewidth]{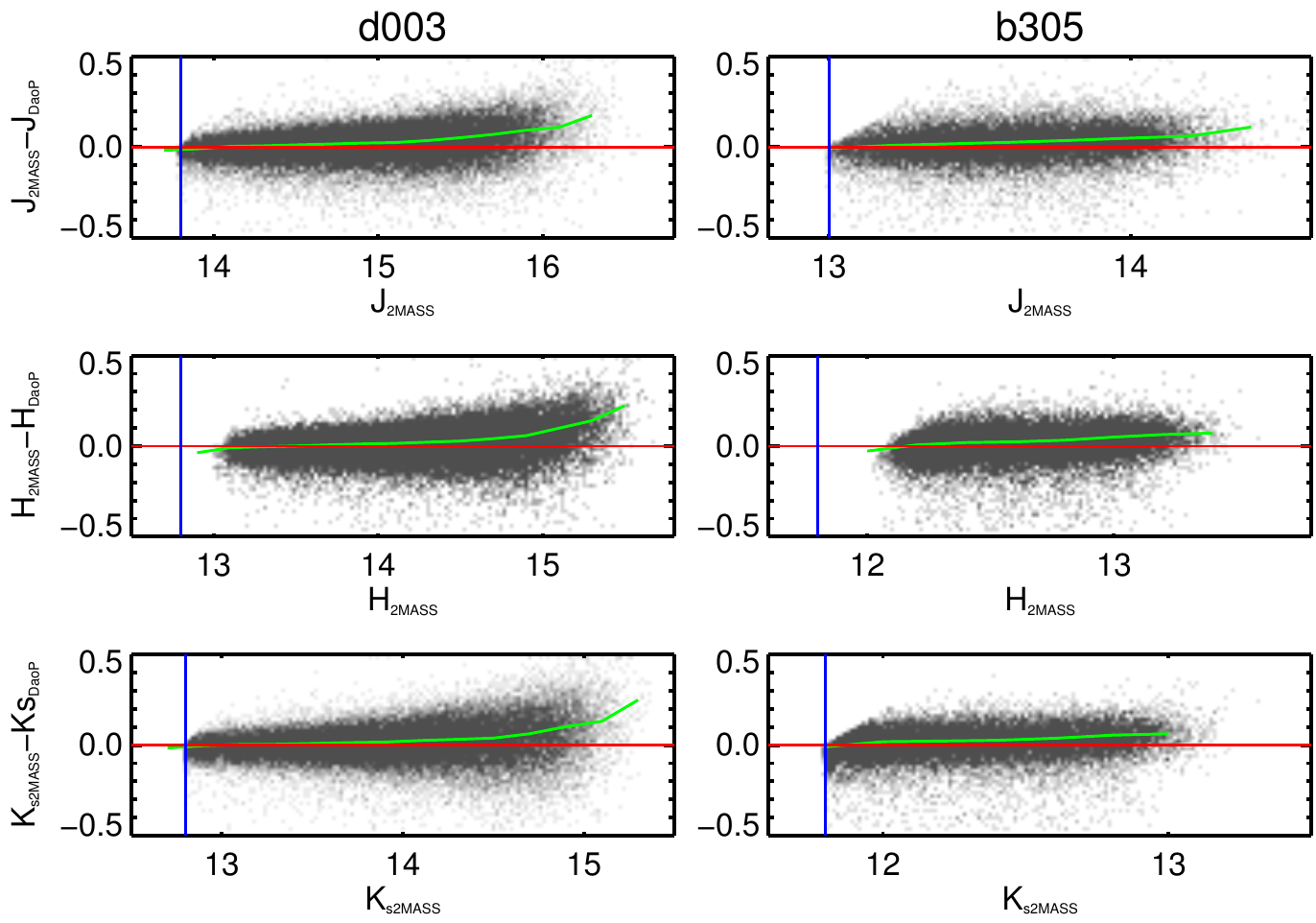}
\caption{2MASS vs. our DaoPHOT photometry in tiles "d003" (\textit{left panels}) and "b305" (\textit{right panels}). The backgrounds show source number density. The green curves show moving medians with a box width of 0.2 mag while the red horizontal lines mark the zero difference value between the 2MASS and our DaoPHOT catalogs. The blue vertical lines show the magnitude limits of ($J$, $H$, $K_s$)=(13.8, 12.8, 12.8) mag and (13.0, 11.8, 11.8) mag for VVV disk tiles and bulge tiles, individually. The source brighter than above limits have been replaced by 2MASS photometric results.}
\label{fig:compare2mass}
\end{figure*}
\end{center}

After the calibration, our DaoPHOT catalog is in the VISTA photometric system. \revile{To date, the 2MASS survey \citep{2mass} is still the main all-sky near-infrared photometric survey. Many models and data in the literature are based on the 2MASS photometric system. Therefore, to enable convenient comparison with previous works in literature, we decided to provide our catalog in both VISTA and 2MASS systems. To make the transfer to the 2MASS system, we re-calculated }
\revile{the transformations ourselves} for each tile using the method described in \citet{soto13}. Figure~\ref{fig:compare2mass} shows the comparison between 2MASS and our DaoPHOT photometric catalog in tiles "d003" and "b305". We also adopt a magnitude limit in the bright end of our DaoPHOT catalog in order to remove the sources that are saturated or affected by residual nonlinearity. This limit is ($J$, $H$, $K_s$) = (13.8, 12.8, 12.8) mag in VVV disk tiles and (13.0, 11.8, 11.8) mag in bulge tiles. The sources brighter than the limit in any band are replaced by 2MASS photometry. \revile{We note that the method suggested by \citet{soto13} does not consider the effect of interstellar extinction. The transformations calculated without the interstellar extinction correction could introduce additional bias \citep{vista-cali}. Therefore, in our catalog, we provide the photometry in both the VISTA and 2MASS systems. The users can use their own transformations to calculate the magnitudes of sources in other photometric systems.}

\subsubsection{Flagging of spurious detections}
\label{subsubsec:removal}

\begin{center}
\begin{figure*}
\includegraphics[width=1.0\linewidth]{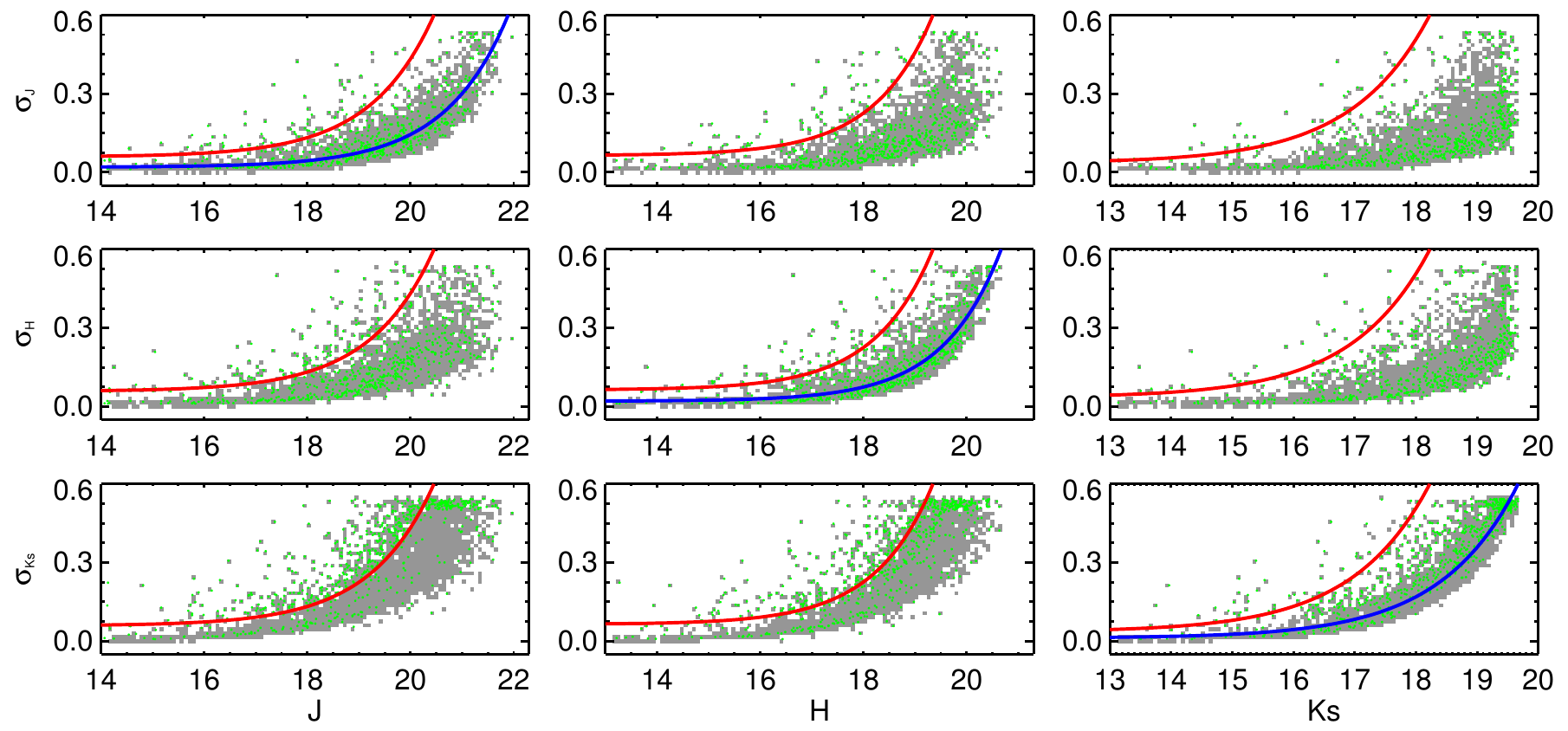}
\caption{Relations between photometric uncertainties and magnitudes for sources in a partial region of "d003" \revitwomz{($\sim$3\arcmin$\times$3\arcmin~central region of tile "d003")}. The gray dots represent all detected sources. The flagged spurious detections are marked with green dots. The blue curves show the exponential fittings for all the sources while the red curves mark the corresponding thresholds of photometric uncertainties that are used to flag the possible spurious detections.}
\label{artifactshow}
\end{figure*}
\end{center}

\begin{center}
\begin{figure}
\includegraphics[width=1.0\linewidth]{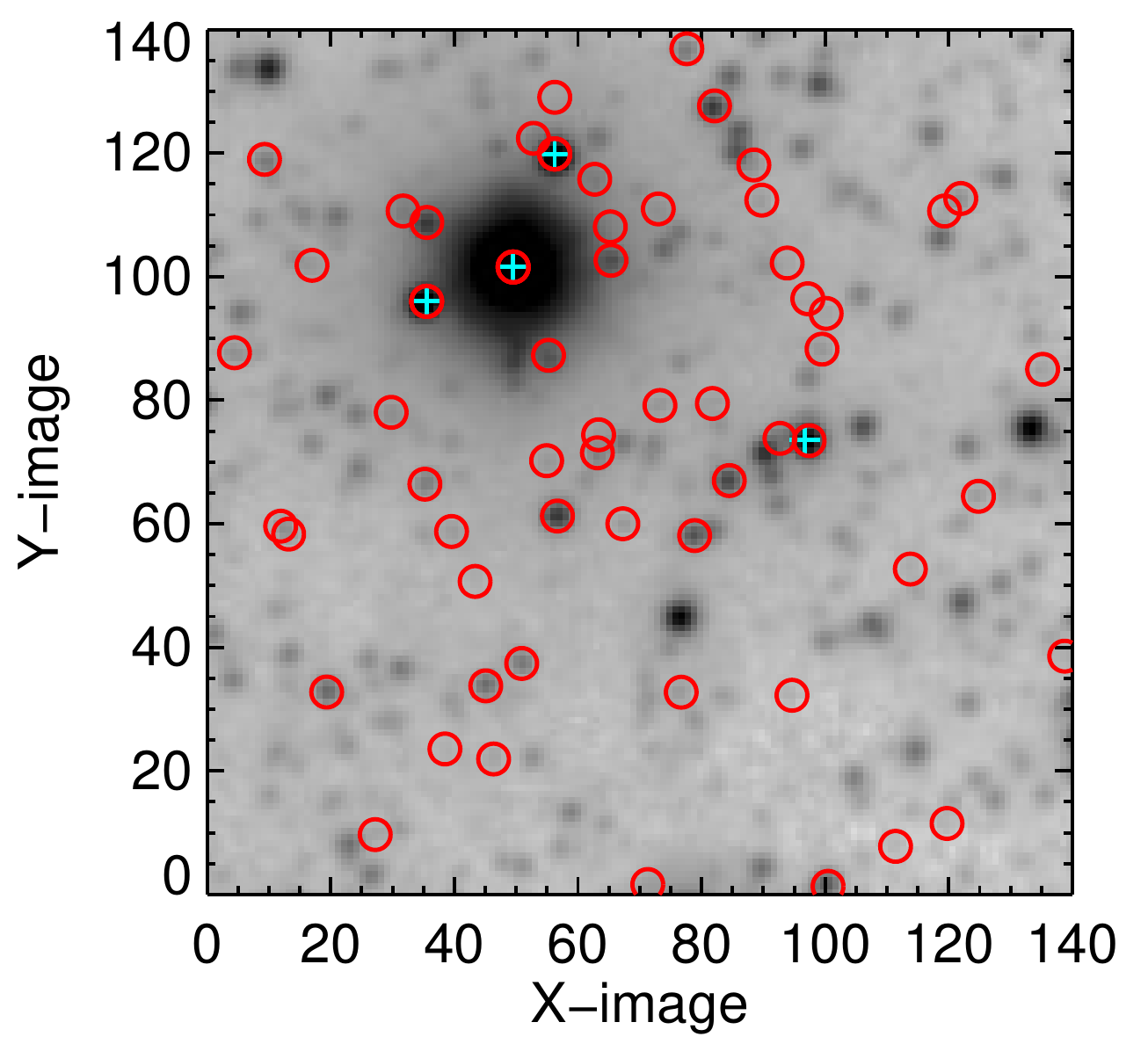}
\caption{Spatial distribution of spurious detections in a partial region \revitwomz{($\sim$50\arcsec$\times$50\arcsec)} of tile "d003". The background is \revionemz{part of a \stasdi~at $K_s$} and the spurious detections are marked with red circles. \revitwomz{We also mark the 2MASS sources that have $J<$13.8\,mag or $H<$12.8\,mag or $K_s<$12.8\,mag with cyan pluses.}}
\label{artifactimgshow}
\end{figure}
\end{center}

Saturated sources usually cause a number of spurious detections around them. To flag the spurious detections, the following two steps are performed. We firstly estimate the mean sky level ($skymode$) for each piece image and define that an image pixel is affected by saturation if its pixel value is higher than a threshold of
\begin{equation*}
skymode+0.8\times(sat-skymode),
\end{equation*}
where $sat$ is the saturation level of the \revionemz{detector} as given in the VIRCAM user manual\footnote{\url{https://www.eso.org/sci/facilities/paranal/instruments/vircam/doc.html}}. We also estimate the mean FWHM of all stars in each piece image and then search pixels around each star within the mean FWHM radius. If there are any surrounding pixels affected by saturation, the star is marked as a source affected by saturation. The second step is the source flagging based on the photometric uncertainties. 
\citet{skzpipeline} suggested that the spurious detections can be rejected with the sigma-clipping method based on the magnitude-error relation. Figure~\ref{artifactshow} illustrates the procedure. For the photometric uncertainties in $m_1$ band and magnitudes in $m_2$ band, we use an exponential function to fit the $\sigma_{m_1}-m_2$ relation:
\begin{equation*}
\sigma_{m1} (m2) = a + e^{b m_2-c}.
\end{equation*}
The corresponding threshold can be obtained as 
\begin{equation*}
\sigma_{m1}^{thres}(m2)=3\sigma_{m1}(m2). 
\end{equation*}
Any source that has a magnitude of $m2$ and an error greater than $\sigma_{m1}^{thres}(m2)$ is flagged as a spurious detection.
For the three bands, that is, $J$, $H$, and $K_s$, there are nine possible combinations. We note that we only fit the relations of $\sigma_J-J$, $\sigma_H-H$, and $\sigma_{K_s}-K_s$ and obtain $\sigma_J^{thres}(J)$, $\sigma_H^{thres}(H)$, and $\sigma_{K_s}^{thres}(K_s)$. For the other combinations we simply assume
\begin{flalign*}
&\sigma_H^{thres}(J)=\sigma_{K_s}^{thres}(J)=\sigma_J^{thres}(J),\\ &\sigma_J^{thres}(H)=\sigma_{K_s}^{thres}(H)=\sigma_H^{thres}(H),\\ &\sigma_J^{thres}(K_s)=\sigma_H^{thres}(K_s)=\sigma_{K_s}^{thres}(K_s).
\end{flalign*}
Figure~\ref{artifactimgshow} shows the spatial distribution of the spurious detections in a partial region of the tile "d003". \revitwomz{We also mark the 2MASS sources that have $J<$13.8\,mag or $H<$12.8\,mag or $K_s<$12.8\,mag. We found that the flagged spurious detections} are mainly the false detections around the bright sources that are affected by saturation or nonlinearity. There are also some very faint sources with large photometric uncertainties. We note that the flagging procedure inevitably flags some real detections as spurious sources. \revitwomz{We found in Fig.~\ref{artifactimgshow} that some real detections that are erroneously flagged as spurious detections are sources with significant differences in the photometric uncertainties among the different bands.} \revionemz{Overall, the process flags $\sim$5$-$27\% of the sources in the
different tiles as spurious detections with a median value of $\sim$10\%.}
\revitwomz{In total, about 73 million sources are flagged as spurious detections.}

\revitwomz{The process above mainly flags sources with relatively large uncertainties. Because we combine the multi-epoch data and perform the photometry on the combined images, we have lost the flux variability of the sources and the variable stars could have large photometric uncertainties. Therefore, the variable stars could be flagged as spurious detections in our DaoPHOT catalog. To look into the magnitude of this effect, we crossmatched the flagged spurious detections with the International Variable Star Index (VSX\footnote{\url{https://www.aavso.org/vsx/}}) catalog \citep{vsx-cat} that is an up-to-date database collecting the known and suspected variable stars from the literature. To account for the large astrometric uncertainties of sources in some of the older publications, we also use a large matching tolerance of 10\arcsec~\citep{drake13}. Finally, 127555 ($<$0.2\%) variable stars are identified in our flagged spurious detections. Of course, the VSX catalog is not a uniform catalog. The future systematic and uniform searches of variables with the VVV data will be helpful to quantify the fraction of variable stars among our flagged spurious detections \citep{vvv-variables}.}




\section{Results and discussion: PSF photometric catalog}
\label{sect:psf-catalog}


With our pipeline, we detect about 926 million sources in the VVV survey area and obtain their photometry. About 342 million sources are in the disk area and about 584 million in the bulge area. We note that we only keep the sources that have detections in all three of the $J$, $H$, and $K_s$ bands. If we exclude the sources flagged as spurious detections (see Sect~\ref{sect:psfphot}), there are about 853 million sources remaining, from which about 319 and 534 million are in the disk and bulge area, respectively. 
As a comparison, the released DoPHOT catalog includes about 180 and 354 million sources with $J, H, K$ detections in the disk and bulge area. 

Our final catalog contains \revile{29} parameters, including the WCS coordinates, photometric magnitudes and uncertainties, and the quality control parameters. An overview of the available columns of the catalog is given in Table~\ref{tab:catalog_para}. 
The provided coordinates ([RAJ2000, DEJ2000]) are in the equatorial system. \revile{The photometric magnitudes ([J\_mag\_2MASS, H\_mag\_2MASS, K$_\textrm{s}$\_mag\_2MASS]) are in the 2MASS system while the magnitudes ([J\_mag\_VISTA, H\_mag\_VISTA, K$_\textrm{s}$\_mag\_VISTA]) are in the VISTA system. All the photometric magnitudes are in the Vega system.} The parameters $N_{\textrm{iter}}$ ([J\_iter, H\_iter, K$_{\textrm{s}}$\_iter]), sharpness ([J\_sharpness, H\_sharpness, K$_{\textrm{s}}$\_sharpness]), and $\chi^2$ ([J\_chi, H\_chi, K$_{\textrm{s}}$\_chi]) describe the PSF-fitting quality. The $N_{\textrm{iter}}$ values give the number of iterations used to fit the stars. The $\chi^2$ refers to the PSF-fitting goodness: $\chi^2=$1 means a perfect fit and a poor fit has a value far from 1. The sharpness is a shape measurement of source detections on the image: sharpness$=$0 means that the detected source is round. The nine spurious detection flags ([JJ\_flag, HH\_flag, K$_s$K$_s$\_flag, JH\_flag, JK$_s$\_flag, HJ\_flag, HK$_s$\_flag, K$_s$J\_flag, $K_s$H\_flag]) in the catalog can be used to filter out the possible spurious detections (see Section~\ref{subsubsec:removal}). In this paper, a spurious detection is defined as a source for which any of the nine flags has a value of 1. Obviously, this is a conservative definition; the potential users can also adopt their own definitions, for example only using [JJ\_flag, HH\_flag, K$_s$K$_s$\_flag] equal 1 to define the spurious detections.

In the following, we first describe the limiting magnitudes and completeness of our catalog and then compare it with the previously released PSF photometric catalog by \citet{vvvpsf}. \revile{We emphasize that the limiting magnitudes and completeness of our catalog are calculated based on the photometric magnitudes in the 2MASS system while the comparison with the DoPHOT catalog is based on the photometric magnitudes in the VISTA system.}

\begin{table}
\scriptsize
\caption{Available data columns of our DaoPHOT photometric catalog for \revionemz{the} VVV survey}
\label{tab:catalog_para}
\centering
\begin{tabular}{llc}
\hline\hline
Column name & Description & unit\\
\hline 
RAJ2000&Right ascension (J2000)& deg\\
DEJ2000 & Declination (J2000) & deg\\
\revile{($J/H/K_s$)\_mag\_VISTA} & \revile{photometric magnitude in VISTA system} & \revile{mag}\\
($J/H/K_s$)\_mag\_2MASS & photometric magnitude in 2MASS system & mag\\
($J/H/K_s$)\_err & photometric error & mag\\
($J/H/K_s$)\_niter & number of PSF-fitting iterations & count\\
($J/H/K_s$)\_sharpness & roundness of object & \\
($J/H/K_s$)\_chi & goodness of PSF fit & \\
($J/H/K_s$)($J/H/K_s$)\_flag & nine spurious detection flags & \revionemz{0 or 1}\\
\hline                                   
\end{tabular}
\end{table}

\subsection{Limiting magnitudes and completeness}

\begin{figure*}
\includegraphics[width=1.0\linewidth]{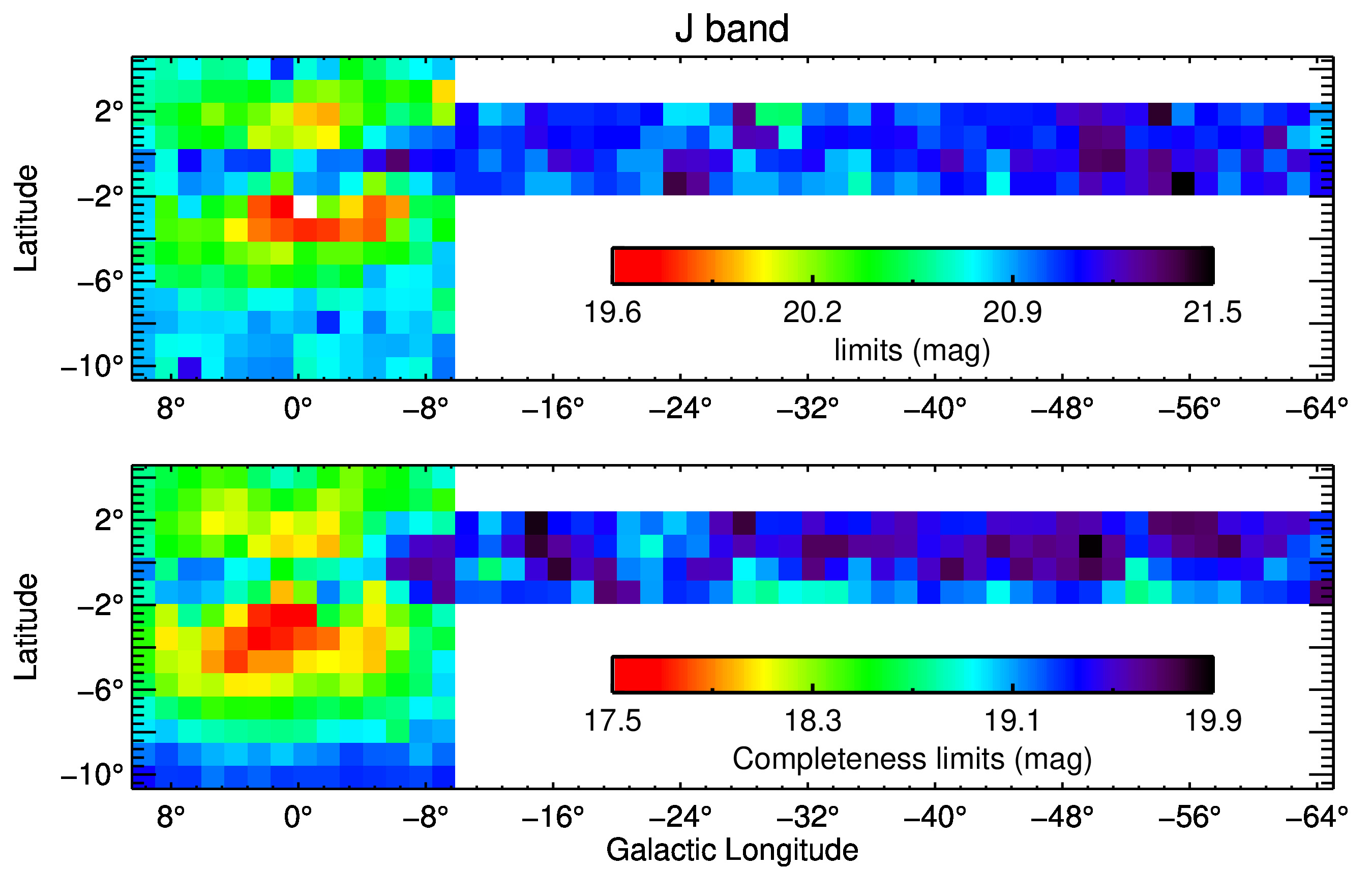}
\caption{5$\sigma$ magnitude limits (\textit{top panel}) and \revionemz{completeness limits} (\textit{bottom panel}) for each VVV tile in the $J$ band.} 
\label{fig:maglimcomj}
\end{figure*}

\begin{figure*}
\includegraphics[width=1.0\linewidth]{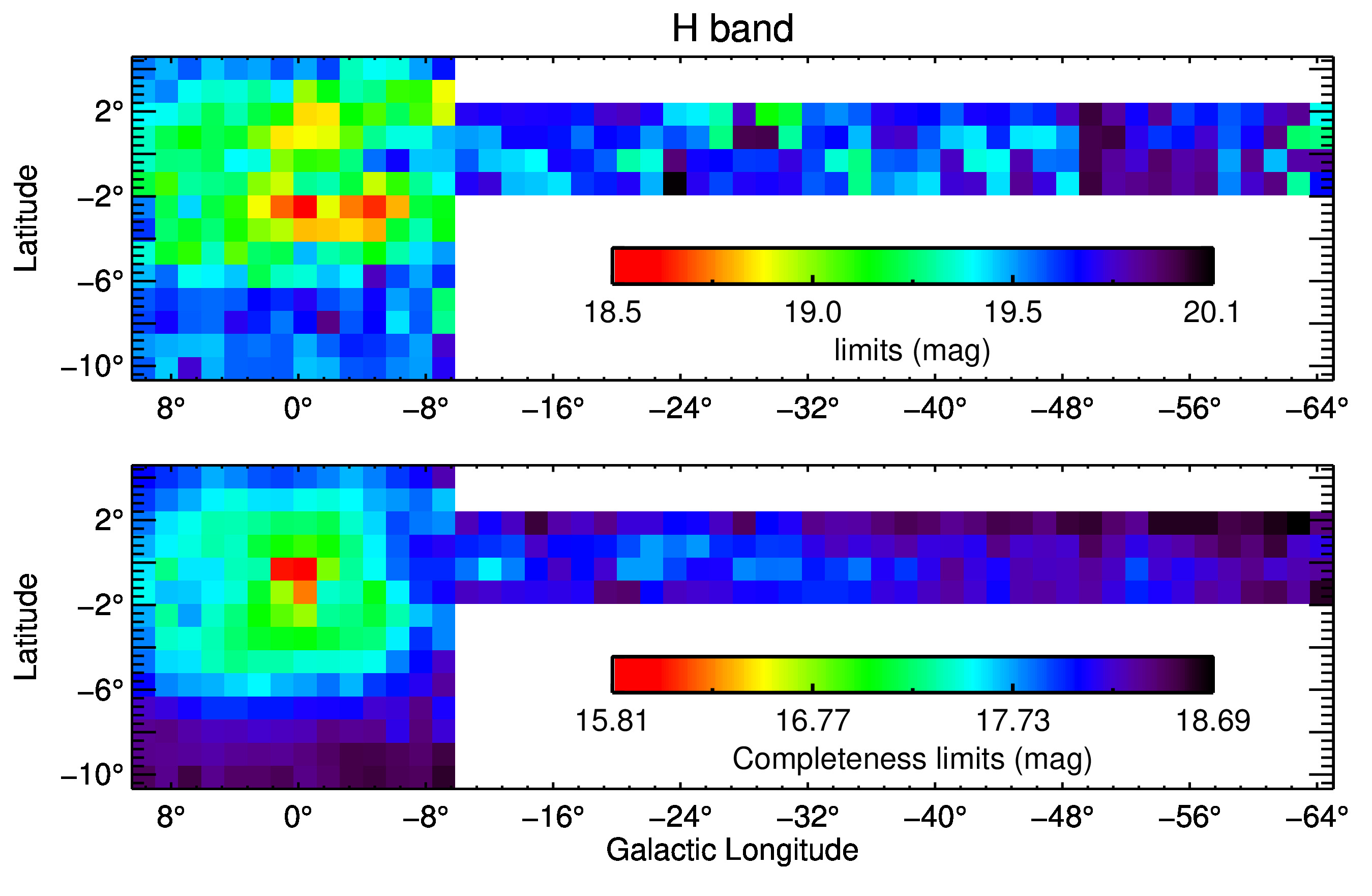}
\caption{5$\sigma$ magnitude limits (\textit{top panel}) and \revionemz{completeness limits} (\textit{bottom panel}) for each VVV tile in the $H$ band.} 
\label{fig:maglimcomh}
\end{figure*}

\begin{figure*}
\includegraphics[width=1.0\linewidth]{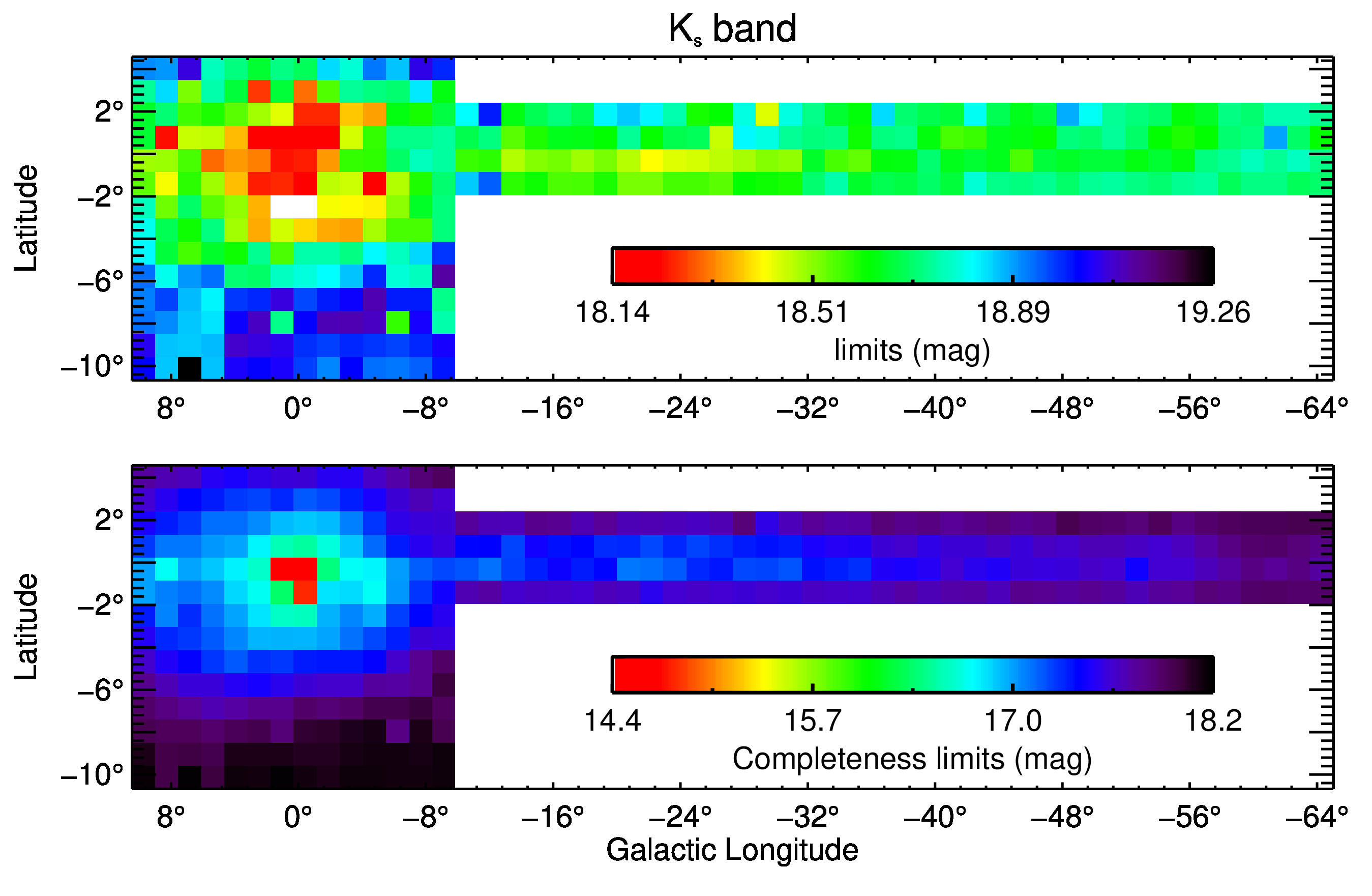}
\caption{5$\sigma$ magnitude limits (\textit{top panel}) and \revionemz{completeness limits} (\textit{bottom panel}) for each VVV tile in the $K_s$ band.} 
\label{fig:maglimcomk}
\end{figure*}

\myemph{The 5$\sigma$ limiting magnitudes are estimated using the sources with photometric uncertainties of $\sigma$ $<$~0.2 
mag after excluding spurious detections. We split each tile into many $\sim$10\arcmin$\times$10\arcmin~subregions and adopt the magnitudes of the faintest source with $\sigma<0.2$\,mag as the 5$\sigma$ limiting magnitudes of each subregion. The 5$\sigma$ limiting magnitudes of the tile are then adopted as the median values of  the limiting magnitudes of all subregions.}  Figures~\ref{fig:maglimcomj}-\ref{fig:maglimcomk} (top panels) show the 5$\sigma$ magnitude limits of all VVV tiles. Due to the different exposure time, image quality, and crowding of the fields, the 5$\sigma$ magnitude limits in $J$, $H$, and $K_s$ bands are in the ranges of 19.6-21.5, 18.5-20.1, and 18.1-19.3 mag with median values of 20.8, 19.5, and 18.7 mag, respectively.

To provide a first-order measurement of completeness at each VVV tile, we constructed single-band brightness distributions in $J$, $H$, and $K_s$ bands for each tile using sources \revionemz{after excluding} spurious detections. In practice, we also split each tile into many $\sim$10$^{\prime}\times$10$^{\prime}$ subregions and make histograms of magnitudes with a bin size of 0.2\,mag in each filter band for each subregion. The completeness of each subregion is estimated roughly as the magnitude of the histogram peak. The completeness of each tile is obtained by averaging the completeness values of all subregions.  Figures~\ref{fig:maglimcomj}-\ref{fig:maglimcomk} (bottom panels) show \revionemz{the completeness of} each VVV tile in $J$, $H$, and $K_s$. This results in the completeness limits of our DaoPHOT photometric catalog of $J\sim$17.5-19.9, $H\sim$15.8-18.7, and $K_s\sim$14.4-18.2 mag. For the VVV disk area, the mean value of completeness is $K_s\sim$17.6 mag with a standard deviation of 0.2 mag. However, for the VVV bulge area, the completeness is highly variable. The outer bulge region has the completeness of $K_s\sim$17.7 mag while the completeness of inner bulge region drops from $K_s\sim$17 mag to only 14.4 mag. The mean value of completeness for the whole VVV bulge area is $K_s\sim$ 17.4 mag with a standard deviation of 0.6 mag.

\subsection{Comparison with the released VVV PSF photometry}

In Sect. 3.2.3 we present a comparison of the brightnesses between our DaoPHOT PSF photometry and the DoPHOT photometry released in \citet{vvvpsf}. Here, we additionally compare the photometries by examining the resulting luminosity functions, \revionemz{photometric difference, color-magnitude diagrams (CMDs), and the color excess maps}. We perform the comparison in six $\sim$10$^{\prime}\times$10$^{\prime}$ regions that are located in the disk and bulge areas, cover different Galactic latitudes, and thus represent regions with different stellar fields and source number densities. 

\subsubsection{Luminosity distribution}\label{sect:lf}

\begin{figure*}
\includegraphics[width=1.0\linewidth]{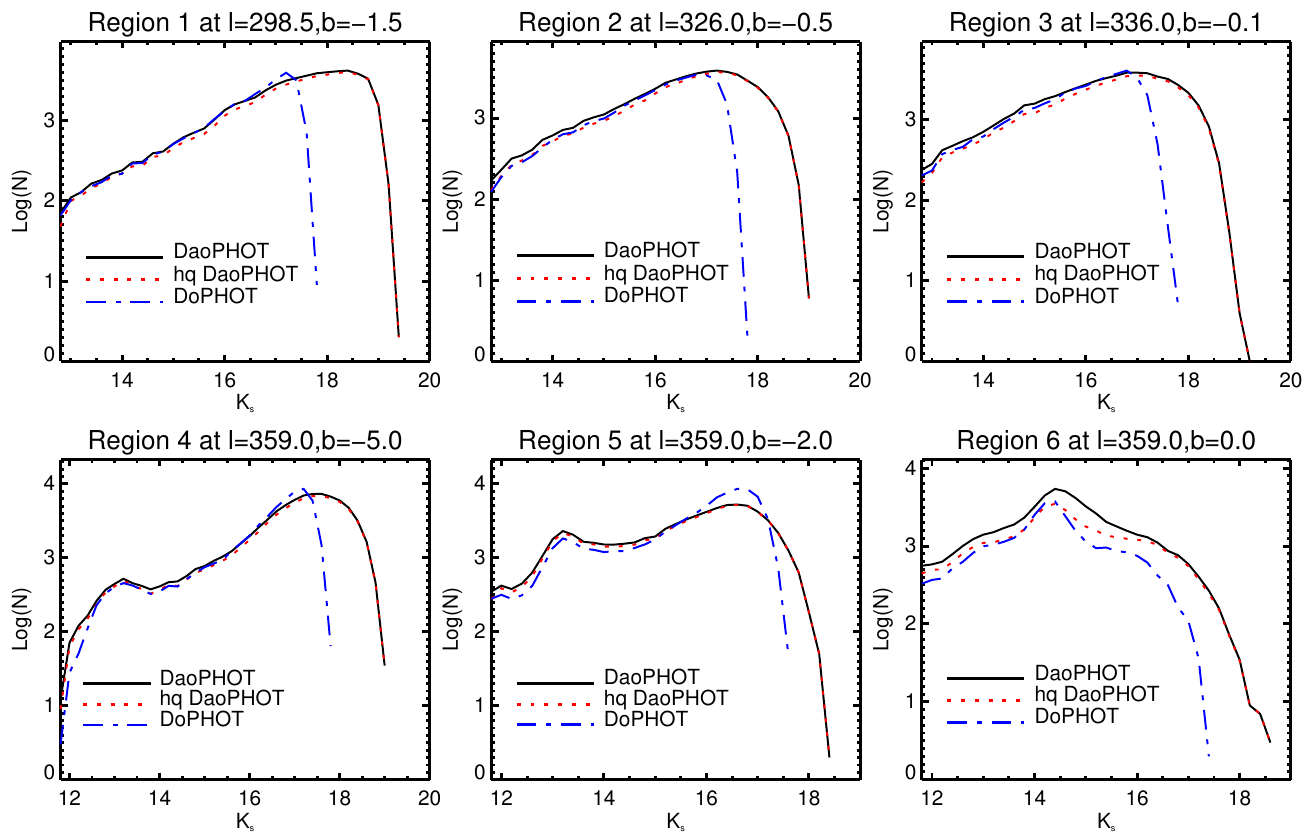}
\caption{Luminosity distributions in the $K_s$ band in six $\sim$10$^{\prime}\times$10$^{\prime}$ regions. All detected DaoPHOT sources are shown with black solid lines while the \revile{hq} 
DaoPHOT sources after removing possible spurious detections are labeled with red dotted lines. The DoPHOT sources retrieved from the released PSF photometric catalog \citep{vvvpsf} are marked with blue dash-dotted lines.}
\label{fig:khist-subregions}
\end{figure*}

\begin{figure*}
\includegraphics[width=\linewidth]{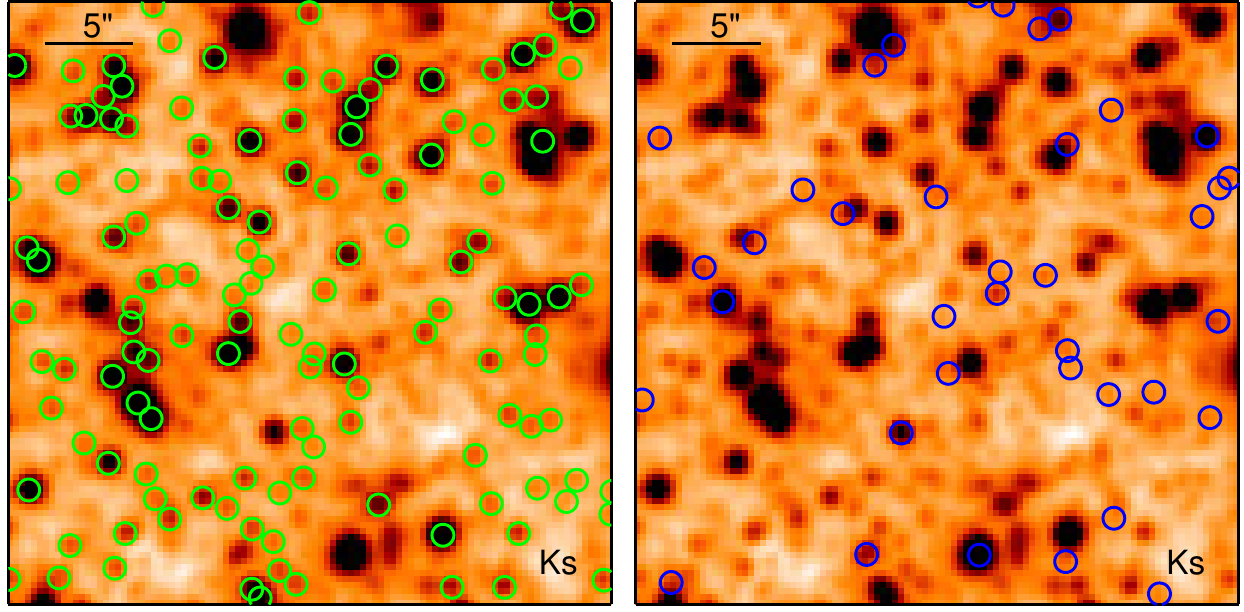}
\caption{Distributions of sources that are detected in both DaoPHOT and DoPHOT catalogs (green circles in \textit{left panel}) and the sources detected only in our DaoPHOT catalog (blue circles in \textit{right panel}) in one test field, region 5. The backgrounds are one partial region \revitwomz{($\sim$35\arcsec$\times$35\arcsec)} of the \stasdis~at $K_s$.}
\label{fig:sourcedetected-region5}
\end{figure*}

Because we performed DaoPHOT photometry on the \revionemz{\stasdis} that have longer total exposure times, we expect to detect more faint sources than is included in the released DoPHOT photometric catalogs. Figure~\ref{fig:khist-subregions} shows the $K_s$ band luminosity distributions for the six test fields. We show the $K_s$ magnitude distributions of all detected DaoPHOT sources and of the high-quality \revile{(hq)} DaoPHOT sources after excluding \revionemz{the spurious detections} (see Sect.~\ref{sect:psfphot}). Our DaoPHOT photometric catalog includes more faint sources and reaches $\sim$1 mag deeper than the released DoPHOT catalog. As an example, Fig.~\ref{fig:sourcedetected-region5} shows the sources detected both in our DaoPHOT catalog and the released DoPHOT catalog in a small region of a \stasdi~in one test field (region 5). The figure also shows the DaoPHOT-only sources, that is, those detected only in our DaoPHOT catalog.

\begin{figure*}
\includegraphics[width=\linewidth]{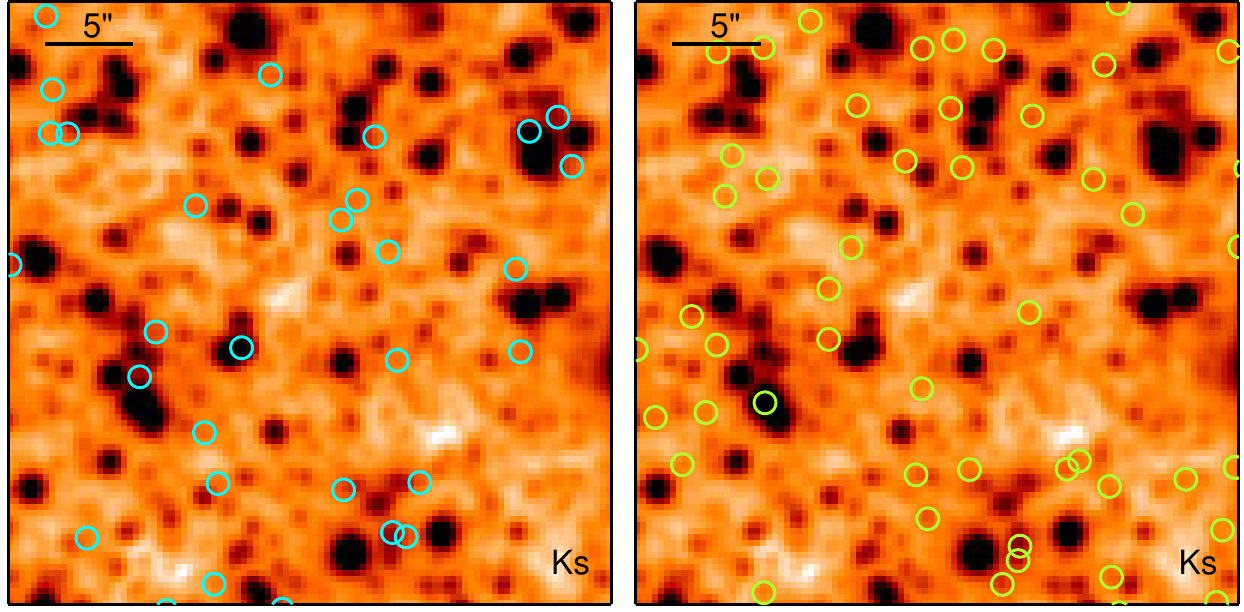}
\caption{Distribution of some DoPHOT-only sources \revionemz{marked with cyan circles} that are neither in our DaoPHOT source catalog nor in the DaoPHOT rejected-source catalog \revitwomz{(\textit{left panel}) and DoPHOT-only sources marked with yellow green circles that are rejected by the DaoPHOT algorithm (\textit{right panel})} on $K_s$ \revionemz{\stasdi~} in \revitwomz{the partial region ($\sim$35\arcsec$\times$35\arcsec) of test field 5.}}
\label{fig:spshow-region5}
\end{figure*}

We note that in some regions in the magnitude range of $K_s\sim$ 16-17 mag the DoPHOT catalog includes more sources than our DaoPHOT catalog (e.g., region 1, 3, 4, and 5). To look further into this, we cross-matched the released DoPHOT catalog with our DaoPHOT catalog in each test field and identified "DoPHOT-only" sources. We found that most of the DoPHOT-only sources are rejected DaoPHOT sources which failed to be fitted by the DaoPHOT PSF models. For example, in region 5 there are 76 999 DoPHOT and 75 155 DaoPHOT sources, and 21 460 (28\%) DoPHOT sources have no DaoPHOT counterparts. Of these 21 460 DoPHOT-only sources, 15 241 (71\%) sources are in the DaoPHOT rejected-source catalog and are therefore removed as spurious detections. \revitwomz{Figure~\ref{fig:spshow-region5} (right panel) shows the distribution of some of these 15241 sources on the $K_s$ \stasdi.} However, there are 6219 (8\%) DoPHOT sources that are neither in the DaoPHOT source catalog nor in the DaoPHOT rejected-source catalog. We examined these sources visually on the images. Figure~\ref{fig:spshow-region5} \revitwomz{(left panel)} shows some of the 6219 sources on the $K_s$ \revionemz{\stasdi}. These sources appear to be very faint and/or highly blended with other sources. Therefore, we suspect that most of them are also spurious detections. These discrepancies could be due to the different algorithms used in DoPHOT and DaoPHOT, as pointed out by \citet{dophot}. DaoPHOT detects objects using a threshold of signal-to-noise ratio (S/N)
and then applies PSF-fitting to the detected objects. During the PSF-fitting process, some objects could fail the fitting and be rejected. In contrast, DoPHOT detects objects using a strict \revile{S/N} threshold and fits all detected objects without rejection. Therefore, if a low \revile{S/N} threshold is adopted, DaoPHOT is likely to detect objects without a well-defined S/N limit, but DoPHOT is inclined to include more spurious detections. We conclude that our DaoPHOT photometric catalog can reach about 1 mag deeper and includes less spurious detections than the released DoPHOT catalog.

\subsubsection{Photometric differences}\label{sect:colordiff}

In section~\ref{sect:photcali}, we compared the photometric magnitudes between our DaoPHOT catalog and the released DoPHOT catalog and found a systematic difference at the faint end. In this section, we investigate these differences further in six test fields that represent different crowding conditions. We emphasize that in this and the subsequent sections (Sect~\ref{sect:cmd} and \ref{sect:colorexcess}), we only consider DoPHOT sources that have detections at $J$, $H$, and $K_s$ simultaneously and DaoPHOT sources that are not spurious detections (see Sect~\ref{subsubsec:removal}). We also require that the uncertainties of the DoPHOT and DaoPHOT sources be 
less than 0.35 mag in the $J$, $H$, and $K_s$ bands.

\begin{figure*}
\includegraphics[width=\linewidth]{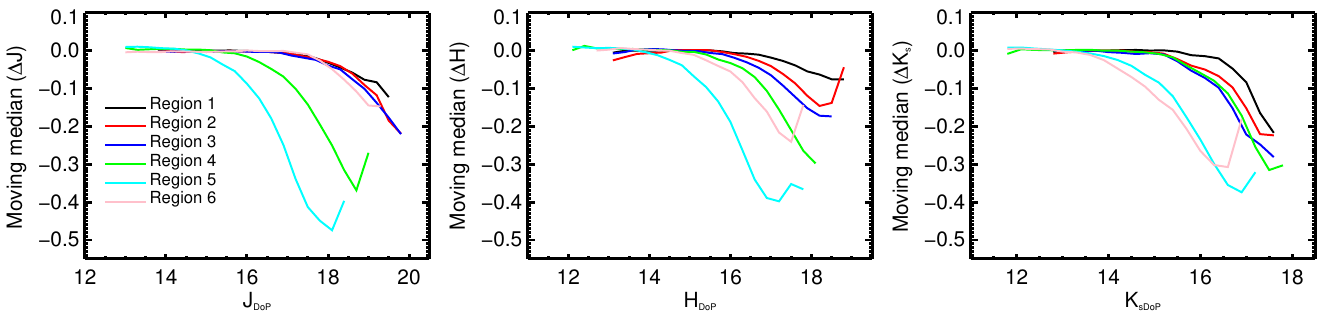}
\caption{Moving medians of the photometric magnitude difference of $\Delta J$ (\textit{left panel}), $\Delta H$ (\textit{middle panel}), and $\Delta K_s$ (\textit{right panel}) between DaoPHOT and DoPHOT catalogs in six $\sim$10$^{\prime}\times$10$^{\prime}$ test fields. The different regions are shown with different color curves.}
\label{fig:magdiff}
\end{figure*}

{\revitwomz{We checked the photometric difference between DaoPHOT and DoPHOT source in \revitwomz{six} test fields and found that} there are obvious systematic differences at the faint ends in all six test fields. To quantify the systematic differences,  Fig.~\ref{fig:magdiff} shows the moving medians of $\Delta J$, $\Delta H$, and $\Delta K_s$ between DaoPHOT and DoPHOT sources as functions of $J$, $H$, and $K_s$ magnitudes, respectively. In the Galactic disk area (regions 1, 2, and 3), the systematic differences appear at $J\sim$17, $H\sim$16, and $K_s\sim$15.5\,mag and can reach up to $\Delta J\sim$0.2, $\Delta H\sim$0.2, and $\Delta K_s\sim$0.3\,mag. In region 4, which is located in the outer bulge, the systematic differences appear at $J\sim$16, $H\sim$15.5, and $K_s\sim$15\,mag and reach up to $\Delta J\sim$0.4, $\Delta H\sim$0.3, and $\Delta K_s\sim$0.3\,mag at $J\sim$18.5, $H\sim$18, and $K_s\sim$17.5\,mag, respectively. In region 5, located in the inner bulge, the systematic differences appear at $J\sim$15, $H\sim$14, and $K_s\sim$14\,mag and reach up to $\Delta J\sim$0.5, $\Delta H\sim$0.4, and $\Delta K_s\sim$0.4\,mag at $J\sim$18, $H\sim$17, and $K_s\sim$17\,mag, respectively. In region 6,  located in the Galactic center, the systematic differences appear at $J\sim$17, $H\sim$15, and $K_s\sim$13.5\,mag and reach up to $\Delta J\sim$0.15, $\Delta H\sim$0.25, and $\Delta K_s\sim$0.3\,mag at $J\sim$19, $H\sim$17.5, and $K_s\sim$16.5\,mag, respectively. Therefore, the differences depend on the location in the Galactic disk and bulge, caused mainly by the local crowding conditions. We also note that the source number density is different in different filter bands in one location. For example, the average source number density of region 6 is similar to that of region 5 at $K_s$ band, but significantly smaller than that of region 5 at $J$ band due to the heavy interstellar extinction towards region 6.}

\begin{figure*}
\includegraphics[width=\linewidth]{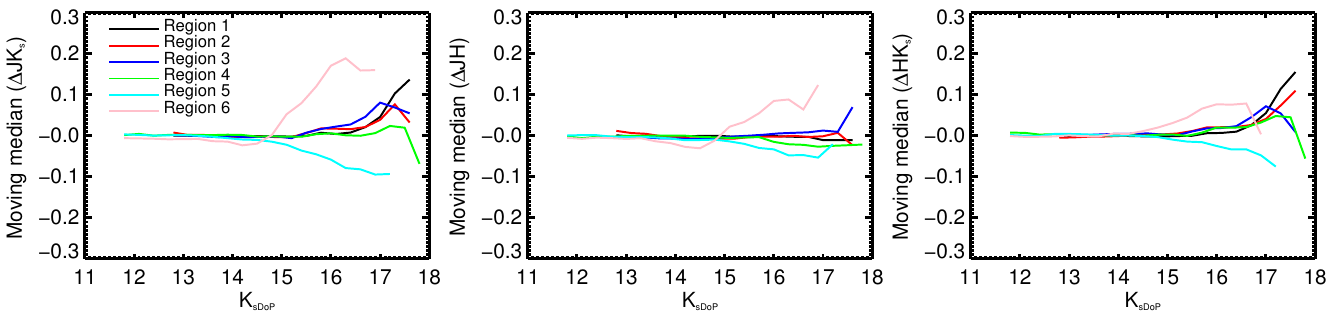}
\caption{Moving medians of the photometric color difference of $\Delta JK_s$ (\textit{left panel}), $\Delta JH$ (\textit{middle panel}), and $\Delta HK_s$ (\textit{right panel}) between DaoPHOT and DoPHOT catalogs as a function of $K_s$ band magnitude for the six $\sim$10$^{\prime}\times$10$^{\prime}$ test fields. The different regions are shown with different color curves.}
\label{fig:colordiff}
\end{figure*}

We also investigate the photometric colors between DaoPHOT and DoPHOT catalogs. 
There are systematic color differences at the faint end, especially in the crowded fields such as regions 5 and 6 which are located close to the Galactic center.
Figure~\ref{fig:colordiff} shows the moving medians of $\Delta JK_s$, $\Delta JH$, and $\Delta HK_s$ between DoPHOT and DaoPHOT catalogs as functions of $K_s$ magnitudes in six test fields. In regions 1, 2, and 3, 
there are no significant differences in the $[J-H]$ color. However, there are differences in $[H-K_s]$ and $[J-K_s]$ colors that appear at $K_s\sim$16\,mag and can reach up to $\sim$0.1\,mag at $K_s\sim$17.5\,mag. In region 4, 
there is no systematic difference for any of the three colors. In region 5, 
the differences in all $[J-K_s]$, $[J-H]$, and $[H-K_s]$ appear at $K_s \sim$15\,mag and can reach up to $\sim$0.1, $\sim$0.05, and $\sim$0.05\,mag at $K_s \sim$17\,mag, respectively. In region 6, 
differences in all colors appear at $K_s \sim$14.5\,mag and can reach up to $\sim$0.2, $\sim$0.1, and $\sim$0.1\,mag at $K_s \sim$16-17\,mag, respectively.

Overall, the systematic photometric magnitude difference between our DaoPHOT catalog and the released DoPHOT catalog is different in the different locations of the Galactic disk and bulge, mainly depending on the local source number density. The systematic color difference is less than 0.1\,mag and appears at $K_s \sim$16\,mag in the Galactic disk area; it reaches up to 0.2\,mag and appears at $K_s \sim$14-15\,mag in the bulge area. In general, the systematic difference of colors between DaoPHOT and DoPHOT catalogs is smaller than that of brightnesses.


\subsubsection{Detectable features in the CMDs}\label{sect:cmd}

\begin{figure*}
\includegraphics[width=\linewidth]{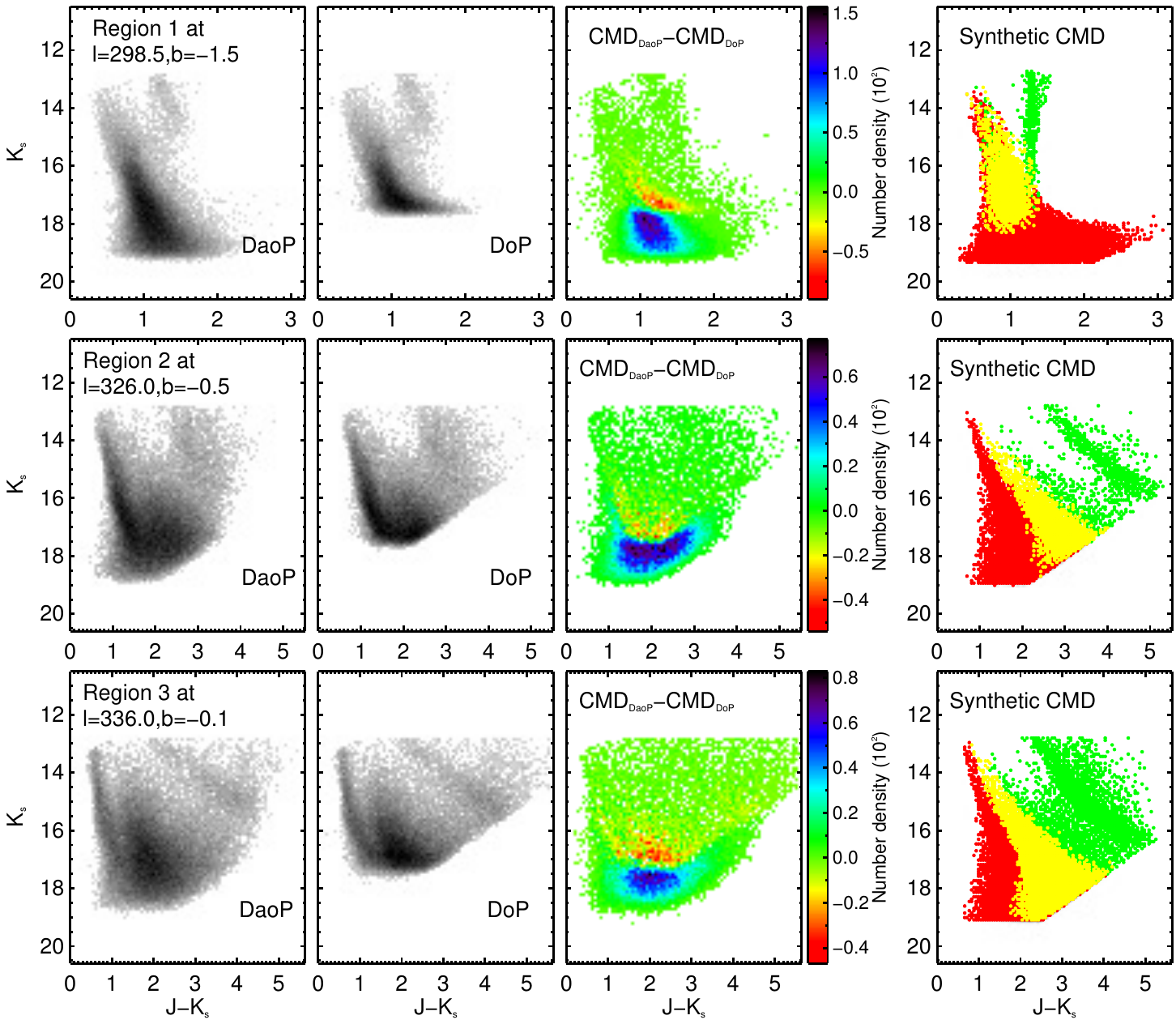}
\caption{$J-K_s$ vs. $K_s$ CMDs of three $\sim$10$^{\prime}\times$10$^{\prime}$ test fields that are located in the Galactic disk area. The \textit{upper, middle, and bottom panels} represent regions 1, 2, and 3, respectively. The center of each field is also marked in the corresponding panel. \textit{Left panels}: CMDs constructed with all DaoPHOT sources; \textit{middle left panels}: CMDs constructed with all DoPHOT sources; \textit{middle right panels}: differential CMDs obtained by subtracting the number density CMD map constructed with DoPHOT catalog from that built with DaoPHOT catalog; \textit{right panels}: the synthetic CMDs built with the Besan\c{c}on Galactic model \citep{robin03}. The MS stars, giants, and subgiants in the Galactic disk are labeled with red, green, and yellow dots, respectively.}
\label{fig:cmd-disk}
\end{figure*}

\begin{figure*}
\includegraphics[width=\linewidth]{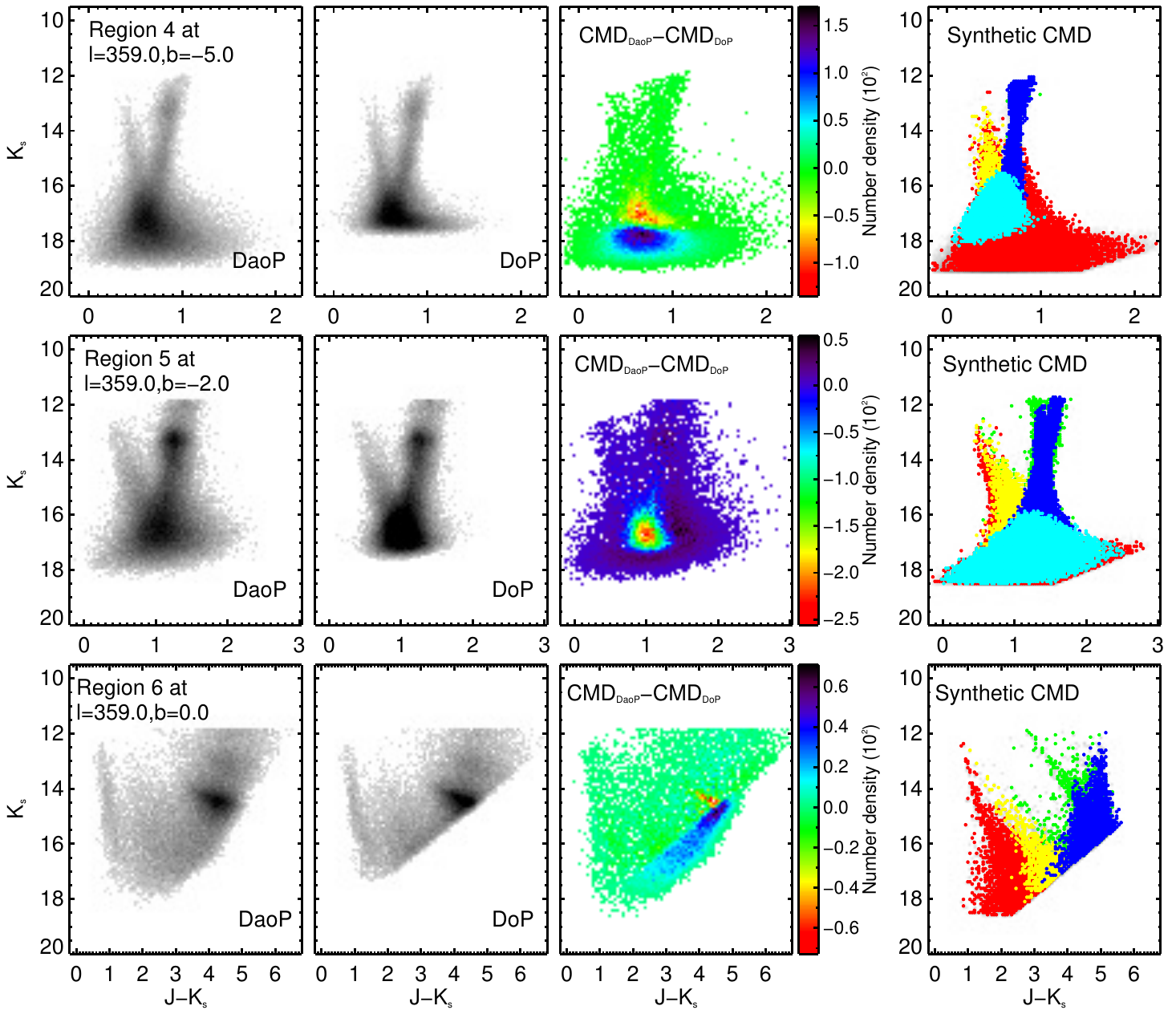}
\caption{$J-K_s$ vs. $K_s$ CMDs in regions 4, 5, and 6 that are located in the Galactic bulge area. The giants and subgiants in the Galactic bulge are marked with blue and cyan dots in the \textit{right panels}. Others are the same as in Fig.~\ref{fig:cmd-disk}.}
\label{fig:cmd-bulge}
\end{figure*}

\revitwomz{The} CMD is a powerful tool to identify the stellar populations in the Galactic disk and bulge \citep{saito12,soto13,vvvpsf}. Figures~\ref{fig:cmd-disk} and \ref{fig:cmd-bulge} compare the CMDs constructed with our DaoPHOT catalog and the released DoPHOT catalog in six test fields. We can see obvious features in the CMDs. To understand the stellar populations corresponding to these detectable features, we obtained synthetic CMDs for the fields using the Besan\c{c}on Galactic model \citep{robin03}. We only show the most populated sequences with different color symbols, that is, the main sequence (MS) disk stars and the giants and subgiants in the disk and bulge.

Figure~\ref{fig:cmd-disk} shows the CMDs in regions 1, 2, and 3. 
In region 1, there are two main features in the CMD. The feature that extends from $K_s\sim$~13 and $[J-K_s]\sim$~0.3 down to $K_s\sim$~19.5 and $[J-K_s]\sim$~2 corresponds to the disk MS and disk subgiant population. The other feature that occupies the color space of $[J-K_s]\sim$~1.4 and $K_s\sim$~14 corresponds to the disk giants. In region 2 and 3, there are three main features. The features with relative blue ($[J-K_s]\sim$~1-2) and red ($[J-K_s]\sim$~3-5) colors correspond to disk MS and disk giants, respectively. There is also another feature between them, extending from $[J-K_s]\sim$~1.5 and $K_s\sim$~14 down to $[J-K_s]\sim$~4 and $K_s\sim$~18. We note that most of disk subgiants are located in this region in the synthetic CMDs.

Figure~\ref{fig:cmd-bulge} shows the CMDs for regions 4, 5, and 6. 
The CMDs of regions 4 and 5 have similar patterns. The different stellar populations are mixed up in the faint end, but there are two distinct branches in the bright end. The blue branch that extends from $[J-K_s]\sim$~0.3 and $K_s\sim$~13 down to $[J-K_s]\sim$~0.7-1 and $K_s\sim$~16 mainly corresponds to the disk MS stars. The red branch \revitwomz{that extends from $[J-K_s]\sim$~1-1.5 and $K_s\sim$~12 down to $[J-K_s]\sim$~0.8-1.2 and $K_s\sim$~16} consists of \revile{disk giants ($\lesssim$20\%) and bulge giants ($\gtrsim$70\%).}
When investigating the CMD of region 6, 
the features become significantly blurred, which is due to the heavy interstellar extinction. Compared with the synthetic CMD, it seems that the different stellar populations such as disk MS stars, disk subgiants, and bulge giants can be roughly separated from one another. There is also an overdensity in the CMD of region 6 at $[J-K_s]\sim$~4.5, $K_s\sim$~14.5. Based on the synthetic CMD it mainly consists of disk giants ($\sim$20\%) and bulge giants ($\sim$80\%).

The CMDs constructed with DaoPHOT and DoPHOT are relatively similar. To quantify the differences between them, we show the differential CMDs in Figs.~\ref{fig:cmd-disk} and \ref{fig:cmd-bulge}. These were obtained by subtracting the number density CMD map built with DoPHOT from that built with DaoPHOT. Indeed, because our DaoPHOT photometry \revile{can reach higher limiting magnitudes} 
than the released DoPHOT photometry, there are positive overdensities at the faint ends of the differential CMDs. There are also negative overdensities at $K_s\sim$~16-17\,mag in some differential CMDs, which means that the DoPHOT catalog includes more sources than the DaoPHOT catalog in this range. We already investigated the excess in Section~\ref{sect:lf} and suggested that it results from the DoPHOT catalog including more spurious detections than our DaoPHOT catalog. Except for the above two features, there are no other significant patterns in the differential CMDs. Therefore, there is no significant difference between the CMDs constructed with DaoPHOT and DoPHOT catalogs in the range $K_s\lesssim$16\,mag. 

\subsubsection{Color excess maps}\label{sect:colorexcess}

We plan to use our DaoPHOT catalogs in the future to derive extinction maps for molecular clouds in the Galactic plane. For this, it is important to investigate the difference of extinction maps constructed with DaoPHOT and DoPHOT catalogs. However, performing extinction mapping for molecular clouds is not a straightforward process, especially when clouds are located at distances larger than approximately 1\,kpc. This is mainly due to the contamination by foreground sources that are located between the clouds and the observer \citep[e.g.,][]{kainulainen11alves}. Therefore, in this section we only compare the color excess maps derived with DaoPHOT and DoPHOT catalogs in six $\sim$1$\degr\times$1$\degr$ test fields. To keep the test simple, we do not try to isolate molecular clouds or exclude any foreground contamination. The six $\sim$1$\degr\times$1$\degr$ test fields have the same centers as the  six previously used $\sim$10$^{\prime}\times$10$^{\prime}$ test fields. 

\begin{figure*}
\includegraphics[width=\linewidth]{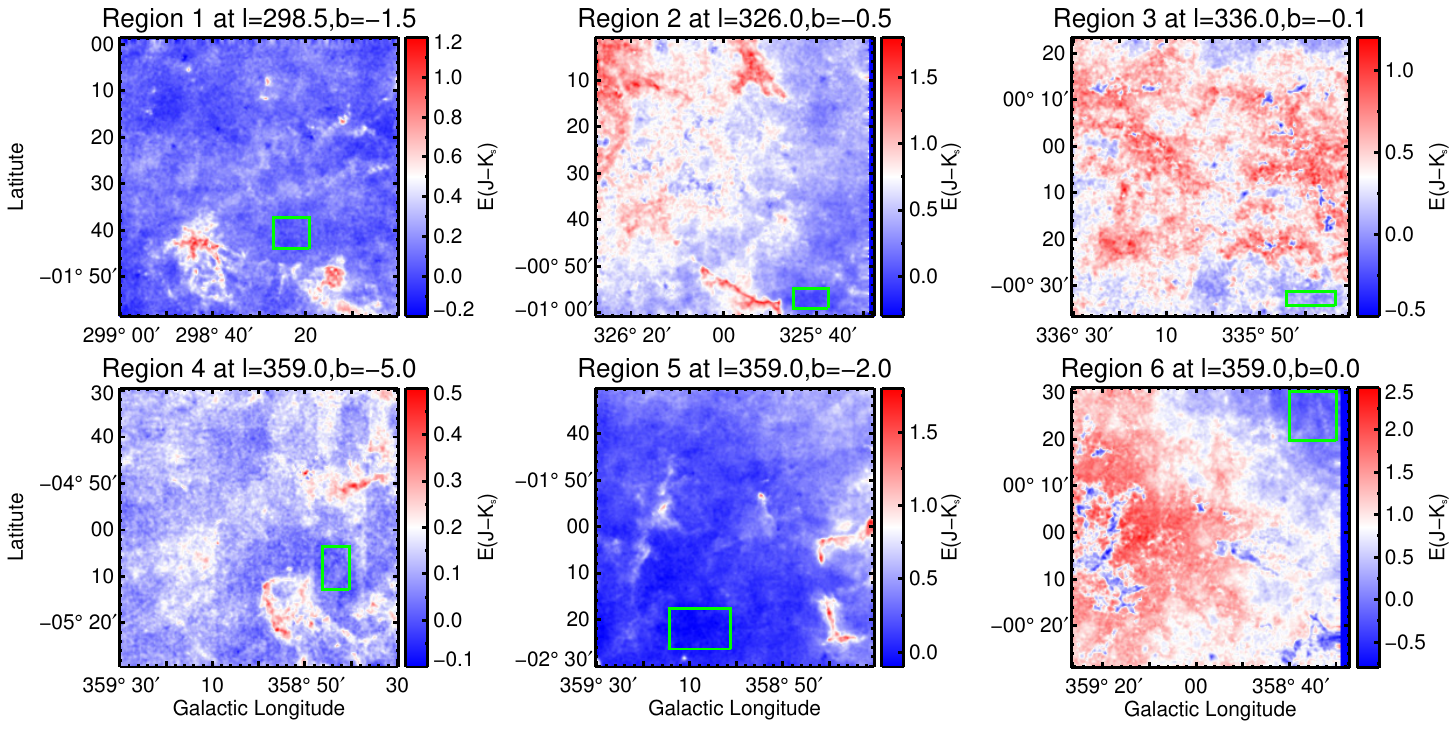}
\caption{Color excess maps of $E(J-K_s)$ in six $\sim$1$\degr\times$1$\degr$ test fields obtained with our DaoPHOT catalog by smoothing the $[J-K_s]$ color excess of sources with a Gaussian beam of FWHM$=$30$^{\prime\prime}$. The green box in each panel shows the control field used for the intrinsic color estimation.}
\label{fig:red-dao}
\end{figure*}

The color excess of each star can be obtained with
\begin{align}
    E(J-K_s)=[J-K_s]_{\textrm{obs}}-[J-K_s]_{\textrm{intrinsic}},
\end{align}
where $[J-K_s]_{\textrm{intrinsic}}$ is the intrinsic color that is usually estimated as the average color of stars in a nearby control field that is free from extinction. In practice, we first adopt an arbitrary value as the intrinsic color and then calculate the color excess of each star. The color excess maps are obtained by smoothing the color excesses with a Gaussian kernel of FWHM$=$30$^{\prime\prime}$ in width. Based on these maps, we select one control field that has relatively low color excess; Figure~\ref{fig:red-dao} shows the locations of the control fields. Second, we use the average $[J-K_s]$ color of the sources in each control field as the intrinsic color and then obtain the final color excess maps. We note that it is very difficult to select the control fields that are free from extinction in the Galactic plane. However, our purpose here is to investigate the difference of the color excess maps built with DaoPHOT and DoPHOT catalogs, not to study extinction itself. Contamination of the control field does not affect the relative values of color excesses derived from the DaoPHOT and DoPHOT catalogs.

\begin{figure*}
\includegraphics[width=\linewidth]{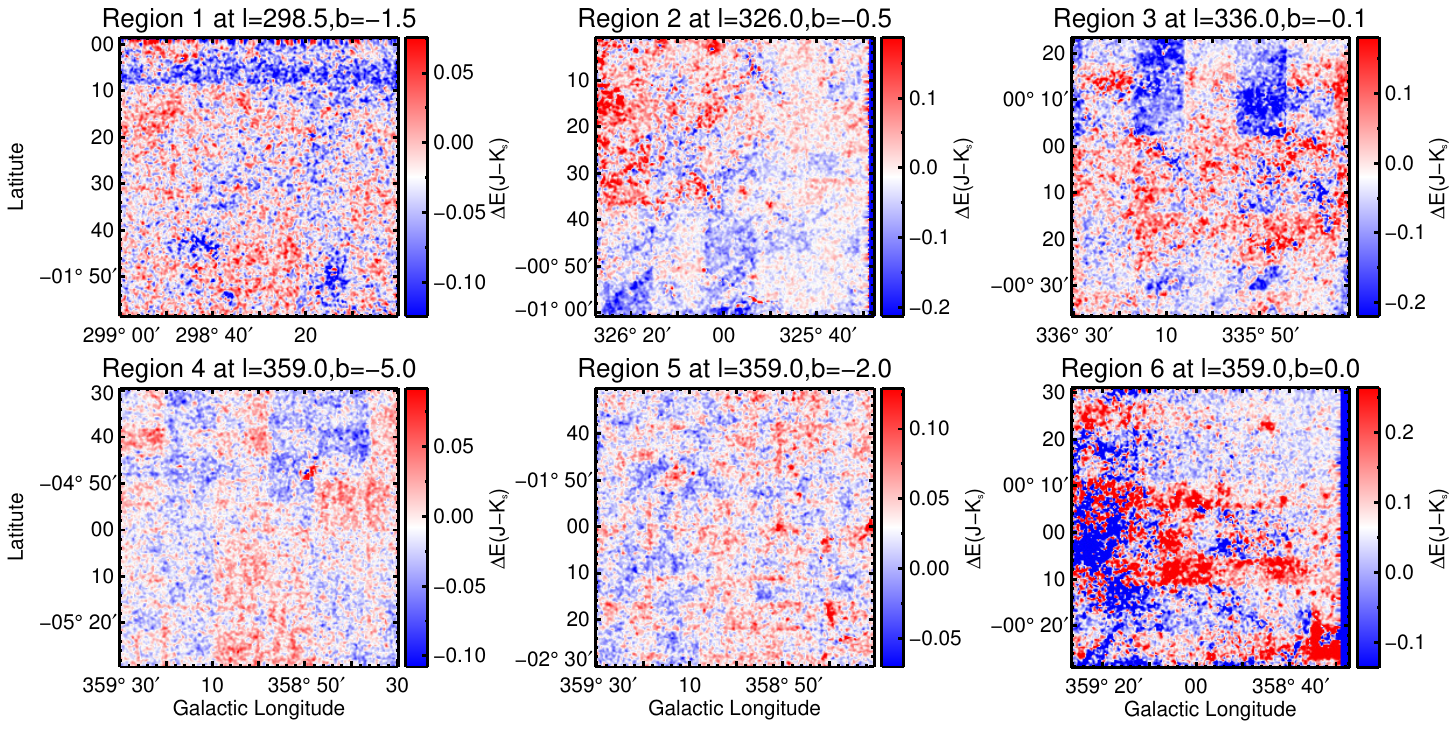}
\caption{\revitwomz{DoPHOT-DaoPHOT} differential maps of color excess, $\Delta E(J-Ks)=$$E_{DoP}(J-K_s)-E_{DaoP}(J-K_s)$, in six $\sim$1$\degr\times$1$\degr$ test fields.}
\label{fig:compare-red}
\end{figure*}

We produce color excess maps using both DaoPHOT and DoPHOT catalogs (see Fig.~\ref{fig:red-dao} for the DaoPHOT maps). To compare the maps, we obtain differential color excess maps by subtracting the DaoPHOT-based map from the DoPHOT-based map. Figure~\ref{fig:compare-red} shows the differential maps. The differential maps show patterns close to the background rms of DaoPHOT and DoPHOT color excess maps. 
The median values and the standard deviations of the six differential maps are -0.02$\pm$0.04, -0.01$\pm$0.07, -0.02$\pm$0.09, -0.01$\pm$0.03, 0.03$\pm$0.03, and 0.06$\pm$0.18. Thus, there is no significant systematic difference between the DaoPHOT and DoPHOT color excess maps for the test fields except region 6 which is one of the most crowded fields in the Galactic bulge. In region 6, the systematic difference between the DaoPHOT and DoPHOT color excess maps is about 0.06, corresponding to the visual extinction of $\sim$0.35\,mag if using the extinction law suggested by \citet{rieke85}.
The above analysis is based on color excess maps made at a spatial resolution of 30$^{\prime\prime}$. We repeated the analysis with maps made at the resolutions of 60$^{\prime\prime}$ and 15$^{\prime\prime}$. The change of spatial resolution does not affect the aforementioned conclusions. 


\subsubsection{Conclusions of the comparisons}

Overall, there are systematic differences in the photometric magnitudes between DaoPHOT and DoPHOT catalogs beginning at $K_s\sim$~14-16\,mag. These differences can reach up to $\Delta K_s\sim$~0.1-0.4\,mag at $K_s\sim$~17\,mag, depending on the crowding of the fields. However, the systematic photometric color differences are less than 0.1\,mag except for the most crowded fields close to the Galactic center. Moreover, there is no significant difference between the color excess maps built with DaoPHOT and DoPHOT sources in the uncrowded fields. In very crowded fields, the difference in the colors of the two catalogs can be up to 0.2\,mag and result in the systematic offset of $\sim$~0.06 ($\Delta A_V\sim$~0.35\,mag) between the color excess maps based on them.

In principle, the DaoPHOT algorithm is more suitable for crowded fields than the DoPHOT algorithm, although neither DaoPHOT nor DoPHOT can recover the true flux of faint stars in crowded fields \citep{dophot,becker07}. Thus, our DaoPHOT catalog should perform better in crowded fields than the released DoPHOT catalog. However, we must emphasize that our DaoPHOT catalog is obtained based on common settings for general photometric parameters; it is difficult to set common parameters for fields with different crowding conditions. Potential users can perform their own PSF photometry using dedicated parameter configurations with our PSF-fitting pipeline should they find that our catalog does not satisfy their scientific requirements. We will release our pipeline to the community in the near future.


\section{Summary and conclusions}\label{sect:conclusions}

In this paper, we present a new PSF photometric catalog based on the DaoPHOT algorithm for the VVV survey data. We also compare the catalog to the PSF photometric catalog recently released by \citet{vvvpsf}. The main results are summarized as follows:

\begin{itemize}
    \item {We developed an automatic PSF-fitting pipeline based on the DaoPHOT algorithm. To reach \revile{the highest possible limiting magnitudes, }
we stack the multi-epoch VVV images and perform photometry on the stacked images. The pipeline mainly uses the PyRAF package and is adapted to run in a multi-core mode. We expect to use the pipeline for other survey data in the future. }
    
    \item Our catalog contains about 926 million sources with detections at $J,~H,$ and $K_s$ bands. About \revionemz{10\%} of the sources are flagged as possible spurious detections. Based on the sources that are not flagged, we estimate the 5$\sigma$ limiting magnitudes to be about 20.8, 19.5, and 18.7 at $J,~H,$ and $K_s$, respectively. 
    Using the peak of the brightness distributions as a first-order measure of completeness \revionemz{(after excluding spurious detections)}, we estimate the completeness limits of our catalog to be about 19.0, 18.0, and 17.5 mag in $J,~H,$ and $K_s$ bands, respectively.

    \item The comparison of our DaoPHOT catalog with the previously released DoPHOT catalog suggests that on average our catalog reaches about 1 mag deeper \revionemz{than the DoPHOT catalog}. \myemph{This results from exploiting different data-reduction and photometric \revitwomz{algorithms}.}  We also find that our DaoPHOT photometry includes less spurious detections than the DoPHOT photometry \myemph{due to the existence of rejection during the PSF-fitting process of DaoPHOT algorithm}.

    
    \item A detailed comparison across different locations in the Galactic disk and bulge suggests a systematic difference of photometric magnitudes for the faint sources between the DaoPHOT and DoPHOT catalogs. This non-negligible systematic difference mainly depends on the crowding of fields and likely results from the different methods of sky background estimation in the catalog pipelines. However, except for very crowded fields close to the Galactic center, there is no significant systematic color difference between the catalogs, and thus no significant difference between the color excess maps built with DaoPHOT and DoPHOT sources. 
\end{itemize}
Developing accurate and efficient photometric pipelines is necessary to make use of the approximately 1 billion sources detected by the VVV survey. This is not a trivial task and the properties of the outcome depend on the chosen approach.  
Our deep PSF photometric catalog can be exploited for a multitude of scientific purposes. In the near future, we plan to use it to derive a new dust extinction map for molecular clouds in the Galactic plane and to investigate the internal structure of the clouds.


\begin{acknowledgements}
\revile{We would like to acknowledge J. P. Emerson for very useful comments about the photometric calibration.} This project has received funding from the European Union's Horizon 2020 research and innovation programme under grant agreement No~639459 (PROMISE). This work was supported by the National Natural
Science Foundation of China (grants No. 11503086). 
This research made use of Astropy,\footnote{\url{http://www.astropy.org}} a community-developed core Python package for Astronomy \citep{astropy:2013, astropy:2018}.
PyRAF is a product of the Space Telescope Science Institute, which is operated by AURA for NASA.
\end{acknowledgements}

\bibliographystyle{aa} 
\bibliography{ref}

\end{document}